\documentclass[3p,times,preprint,review,11pt,authoryear]{elsarticle}

\usepackage[utf8]{inputenc} % German umlaut
\usepackage[english]{babel} % English language
\usepackage{textgreek} %GreekLetters
\usepackage[T1]{fontenc}
\usepackage{natbib} % Citation processing
\usepackage{geometry} % Margin settings
\usepackage{amsmath} % The amsthm package provides extended theorem environments
\usepackage{amsfonts}
\usepackage{amssymb} % The amssymb package provides various useful mathematical symbols
\usepackage{siunitx} % SI units with the command e.g. \si{\meter} or \SI{3}{\micro\meter}
\usepackage{graphicx} % Graphic inclusion
\usepackage[disable]{endfloat} % If empty floats are placed at the end of the PDF. If option "disable" the floats are placed where they are placed in the code
\usepackage{booktabs} % Nice layout of tables
\usepackage{multicol} % Adjusting margins for multicolumn and single column output
\usepackage{multirow} % Create tabular cells spanning multiple rows
\usepackage{threeparttable} % tables with captions and notes all the same width
\usepackage{blindtext} % Create blindtext with the command \blindtext
\usepackage{verbatim} % Comment out text in the environment \begin{comment} your text \end{comment}
\usepackage[colorlinks = true, urlcolor=blue, linkcolor=black, citecolor=blue]{hyperref}
\usepackage{xr-hyper}
\usepackage[nameinlink,noabbrev,capitalize]{cleveref}
\usepackage[singlelinecheck = false]{caption} % Use this package for more than one caption per table
\usepackage{xcolor}
\usepackage{booktabs} % nice layout of tables
\usepackage{multicol} % adjusting margins for multicolumn and single column output
\usepackage{multirow} % create tabular cells spanning multiple rows
\usepackage{threeparttable} % tables with captions and notes all the same width
\usepackage{nicefrac}
\usepackage[commandnameprefix=always,commentmarkup=footnote,authormarkup=none,authormarkuptext=name]{changes} % Remove the change markup after acknowledging or rejecting the changes in the PDF using the option [final]

\journal{Earth and Planetary Science Letters}

\begin{document}
%%%%%%%%%%%%%%%%%%%%%%%%%%%%%%%%%%%%%%%%%%%%%%%%%%%%%%%%%%%%%%%%%%%%%%%%%%%%%%%%%%%%%%%%%%%%%%%%%%%%
%%%%%%%%%%%%%%%%%%%%%%%%%%%%%%%%%%%%%%%%%%%%%%%%%%%%%%%%%%%%%%%%%%%%%%%%%%%%%%%%%%%%%%%%%%%%%%%%%%%%
%% Frontmatter

\begin{frontmatter}

%%%%%%%%%%%%%%%%%%%%%%%%%%%%%%%%%%%%%%%%%%%%%%%%%%%%%%%%%%%%%%%%%%%%%%%%%%%%%%%%%%%%%%%%%%%%%%%%%%%%
%% Title, authors and addresses

%% use the tnoteref command within \title for footnotes; e.g.: \title{My first scientific paper\tnoteref{label1,label2}}
%% use the tnotetext command for theassociated footnote; e.g.: \tnotetext[label1]{footnote 1}, \tnotetext[label2]{footnote 2}
%% use the fnref command within \author or \affiliation for footnotes;
%% use the fntext command for theassociated footnote;
%% use the corref command within \author for corresponding author footnotes;
%% use the cortext command for theassociated footnote;
%% use the ead command for the email address,
%% and the form \ead[url] for the home page:
%% \title{Title\tnoteref{label1}}
%% \tnotetext[label1]{}
%% \author{Name\corref{cor1}\fnref{label2}}
%% \ead{email address}
%% \ead[url]{home page}
%% \fntext[label2]{}
%% \cortext[cor1]{}
%% \affiliation{organization={},
%%            addressline={}, 
%%            city={},
%%            postcode={}, 
%%            state={},
%%            country={}}
%% \fntext[label3]{}

%%%%%%%%%%%%%%%%%%%%%%%%%%%%%%%%%%%%%%%%%%%%%%%%%%%%%%%%%%%%%%%%%%%%%%%%%%%%%%%%%%%%%%%%%%%%%%%%%%%%
%% Title of the manuscript
% Limit to 12 words or less than 100 characters; shorter is better.
%\title{Redox state of peridotitic and basaltic glasses: Implications for the redox state evolution of the Earth and Exoplanets}

%\title{The influence of low-spin ferrous iron on the oxidation state of the Earth's mantle}
\title{Low-spin ferrous iron suppresses mantle oxidation beyond Earth-like pressures}

%%%%%%%%%%%%%%%%%%%%%%%%%%%%%%%%%%%%%%%%%%%%%%%%%%%%%%%%%%%%%%%%%%%%%%%%%%%%%%%%%%%%%%%%%%%%%%%%%%%%
%% Authors, affiliations and footnotes
\author[1]{Alice Girani}
%\ead{alice.girani@eaps.ethz.ch}

\author[1]{Sylvain Petitgirard}
%\ead{petitgirardsylvain@gmail.com}

\affiliation[1]{organization={ETH Zürich, Department of Earth and Planetary Sciences},%Department and Organization
            city={Zürich},
            country={Switzerland}}
            
\author[2]{Sergey Yaroslavtsev}
%\ead{sergey.yaroslavtsev@esrf.fr}

\author[2]{Georgios Aprilis}
%\ead{georgios.aprilis@esrf.fr}

\affiliation[2]{organization={European Synchrotron Radiation Facility},
            city={Grenoble},
            country={France}}

\author[3]{James Badro}
%\ead{badro@ipgp.fr}
\affiliation[3]{organization={Université Paris Cité, Institut de Physique du Globe de Paris, CNRS},
            city={Paris},
            country={France}}

\author[4]{Antoine Bézos}
%\ead{antoine.bezos@univ-nantes.fr}
\affiliation[4]{organization={Nantes Université, Univ Angers, Le Mans Université, CNRS, Laboratoire de Planétologie et Géosciences, LPG UMR 6112},
            city={44000 Nantes},
            country={France}}

\author[5,6]{Hugh St. C. O'Neill}
%\ead{hugh.oneill@monash.edu}
\affiliation[5]{organization={School of Earth, Atmosphere and Environment, Monash University, Department of Earth Sciences},
            city={Melbourne},
            country={Australia}}
\affiliation[6]{organization={Guangzhou Institute of Geochemistry, Chinese Academy of Sciences}, city={Guangzhou},
            country={China}}

\author[1]{Paolo A. Sossi \corref{cor1}}
\ead{paolo.sossi@eaps.ethz.ch}
\cortext[cor1]{Corresponding author.}

%%%%%%%%%%%%%%%%%%%%%%%%%%%%%%%%%%%%%%%%%%%%%%%%%%%%%%%%%%%%%%%%%%%%%%%%%%%%%%%%%%%%%%%%%%%%%%%%%%%%
%% Abstract 
%\linenumbers
\begin{abstract}

\noindent The Earth's mantle has elevated Fe$^{3+}$ relative to those of other rocky bodies, a property thought to reflect the disproportionation of ferrous iron into its metallic and ferric counterparts during core formation at elevated pressures. However, whether more massive planets (`super-Earths') become increasingly oxidised is poorly known for lack of knowledge of the oxidation- and electronic state of iron at extreme pressures. We present \textit{in-situ} energy-domain synchrotron Mössbauer spectra of $^{57}$Fe-enriched peridotitic- and basaltic glasses at 298~K compressed from 1~bar to 174~GPa in a diamond anvil cell. Three glasses were synthesised with different Fe$^{3+}$/[Fe$^{3+}$~+~Fe$^{2+}$] ratios; 0.02~$\pm$~0.02 (Fe$^{2+}$-basaltic, peridotitic) and 1.00~$\pm$~0.02 (Fe$^{3+}$-basaltic), respectively, as determined by colourimetry. 
While the spectrum of pure Fe$^{3+}$-basaltic glass shows minimal changes in its hyperfine parameters up to 174~GPa, the spectra of Fe$^{2+}$-peridotitic and basaltic glasses are fit by two doublets, D$_1$ and D$_2$. At 1 bar, their relative intensities are $\sim$92~\% and $\sim$8~\%, respectively, but the integral area ratio, D$_2$/(D$_1$~+~D$_2$), reaches 0.65 by 172 GPa.
Because this transition is reversible with pressure and no metallic iron is detected, the D$_2$ feature is ascribed to Fe$^{2+}$ in its low spin (LS) state, whereas D$_1$ is consistent with Fe$^{2+}$ high spin (HS). 
Consequently, the Fe$^{3+}$/[Fe$^{3+}$+Fe$^{2+}$] of planetary mantles at constant relative \textit{f}O$_2$ increase to a maximum near $\sim$40~GPa, before decreasing at higher pressures due to the stabilisation of Fe$^{2+}_{\mathrm{{LS}}}$. This peak coincides with estimated core-mantle equilibrium on Earth, implying that its uniquely oxidised mantle and habitable state may result from core formation within a Goldilocks pressure range. Secondary atmospheres are predicted to transition from H$_2$-rich for Moon-sized bodies, to CO-rich for Earth-like planets and H$_2$- and CH$_4$-bearing around super-Earths and sub-Neptunes.

\end{abstract}

%%%%%%%%%%%%%%%%%%%%%%%%%%%%%%%%%%%%%%%%%%%%%%%%%%%%%%%%%%%%%%%%%%%%%%%%%%%%%%%%%%%%%%%%%%%%%%%%%%%%
%% Research highlights environment
% Highlights should be submitted in a separate editable file, named Highlights" in the online submission system.
% 3-5 bullet points of max. 85 characters, including spaces, per bullet point.
% Highlight bullet points capture the novel results of the research done for this study as well as new methods that were used during the study (if any).
\begin{comment}
\begin{highlights}
\item \red{Research highlight 1}
\item \red{Research highlight 2}
\item \red{Research highlight 3}
\item \red{(Research highlight 4)}
\item \red{(Research highlight 5)}
\end{highlights}
\end{comment}

%% Keywords environment
% Provide a max. of 6 keywords, using American spelling
% Categorization: General research area, special research area, method, theory

\end{frontmatter}

%%%%%%%%%%%%%%%%%%%%%%%%%%%%%%%%%%%%%%%%%%%%%%%%%%%%%%%%%%%%%%%%%%%%%%%%%%%%%%%%%%%%%%%%%%%%%%%%%%%%
%%%%%%%%%%%%%%%%%%%%%%%%%%%%%%%%%%%%%%%%%%%%%%%%%%%%%%%%%%%%%%%%%%%%%%%%%%%%%%%%%%%%%%%%%%%%%%%%%%%%
%% EPSL guidelines

% Restricted article length EPSL: 6500 words. It includes headings, citations, and equations within the main text. This excludes abstract, figures, tables, figures and table captions, acknowledgments, references, appendices, and supplementary files.
% Total number of figures and tables is restricted to 10.
% Number of references should not exceed 70.

%%%%%%%%%%%%%%%%%%%%%%%%%%%%%%%%%%%%%%%%%%%%%%%%%%%%%%%%%%%%%%%%%%%%%%%%%%%%%%%%%%%%%%%%%%%%%%%%%%%%
%%%%%%%%%%%%%%%%%%%%%%%%%%%%%%%%%%%%%%%%%%%%%%%%%%%%%%%%%%%%%%%%%%%%%%%%%%%%%%%%%%%%%%%%%%%%%%%%%%%%
%% Main text
%\linenumbers

%%%%%%%%%%%%%%%%%%%%%%%%%%%%%%%%%%%%%%%%%%%%%%%%%%%%%%%%%%%%%%%%%%%%%%%%%%%%%%%%%%%%%%%%%%%%%%%%%%%%
%%%%%%%%%%%%%%%%%%%%%%%%%%%%%%%%%%%%%%%%%%%%%%%%%%%%%%%%%%%%%%%%%%%%%%%%%%%%%%%%%%%%%%%%%%%%%%%%%%%%
%% Introduction

\section{Introduction} \label{introduction}

The Earth, as the only planet known to support life, may owe its importance to the uniquely oxidised character of the present-day atmosphere. The Fe$^{3+}$/[Fe$^{2+}$~+~Fe$^{3+}$] (equivalently, Fe$^{3+}$/$\Sigma$Fe) ratio of the present-day Earth's upper mantle, 0.037 \citep{sossi_redox_2020} is higher than in the mantle of any other terrestrial planet. In a magma ocean, Fe$^{3+}$/$\Sigma$Fe could reach $\sim$~0.09 if electron exchange with Cr is considered \citep{hirschmann_magma_2022}, though pressure would diminish this effect \citep{berry2021}. This Fe$^{3+}$/$\Sigma$Fe ratio would have produced an atmosphere composed predominantly of CO and CO$_2$, had it been in equilibrium with silicate liquid \citep{sossi_redox_2020, gaillard_redox_2022, hirschmann_magma_2022}, plausibly following the Moon-forming impact \citep{canup_origin_2001,caracas_meltcrystal_2019}. However, the origin of such high Fe$^{3+}$/$\Sigma$Fe ratios in the Earth's mantle remains uncertain. Leading hypotheses suggest the stabilisation of ferric iron at high pressures attained during core formation uniquely on Earth may be responsible \citep{frost_redox_2008,armstrong_deep_2019, deng_magma_2020, zhang_ferric_2024}. 
%Because core formation on the Earth is likely the only event capable of internally redistributing mass on a scale sufficient to modify the redox state of Earth's mantle, 
Consequently, the Fe$^{3+}$/$\Sigma$Fe ratios of silicate liquids representative of the mantle, at the pressure-temperature ($P-T$) conditions of equilibrium with the core are of critical importance to scrutinising this hypothesis.\\

\noindent Recent studies \citep{armstrong_deep_2019,deng_magma_2020, hirschmann_magma_2022, kuwahara2023hadean, zhang_ferric_2024} reveal that the Fe$^{3+}$/$\Sigma$Fe ratio in silicate melts increases with increasing pressure beyond $P~>~10$~GPa at constant $f$O$_2$ relative to common mineral buffers.
%the partial molar volume change ($\Delta V$) of Fe$^{3+}$O$_{1.5}$~-~Fe$^{2+}$O is negative beyond $P~>~10$~GPa. 
%As such, FeO$_{\text{1.5}}$ is thought to be more compressible than FeO over the pressure range in the magma ocean relevant to core formation, and the Fe$^{3+}$/$\Sigma$Fe in equilibrium with core-forming alloy at constant oxygen fugacity ($f$O$_2$) increases with pressure.
If the magma ocean were to adiabatically decompress (\textit{e.g.,} via convection) to the surface isochemically (\textit{i.e.,} at constant Fe$^{3+}$/$\Sigma$Fe), then increasingly massive planetary bodies would have more oxidising atmospheres with higher CO$_2$/CO and H$_2$O/H$_2$ ratios, all else being equal \citep{sossi_redox_2020,deng_magma_2020}. However, studies on peridotitic compositions relevant to the Earth's mantle are sparse. \citet{armstrong_deep_2019} and \citet{zhang_ferric_2024} examined broadly andesitic- to basaltic- silicate glasses at 1 bar quenched from up to 71 GPa and $\sim$4300~K. These liquids are, however, depleted in Mg and enriched in Al relative to Earth's primitive mantle. On the other hand, \citet{deng_magma_2020} studied ultramafic compositions in the simplified Mg-Si-Fe-O system by \textit{ab-initio} calculations, but they neglected the possible effects of spin transitions in both Fe$^{2+}$ and Fe$^{3+}$ in their equations of state. Finally, the experiments of \citet{kuwahara2023hadean}, while performed on quenched, broadly peridotitic glasses in equilibrium with iron metal, show evidence for devitrification, which could lead to modification of Fe$^{3+}$/$\Sigma$Fe ratios. \\

Additional uncertainty stems from the difficulty in interpreting Mössbauer spectra of silicate glasses with low ($<$10~\%) Fe$^{3+}$/$\sum$Fe \citep{virgo_structural_1985,cottrell_high-precision_2009,berry_re-assessment_2018}. Because magmatic processes on the Earth and planets occur below the fayalite-magnetite-quartz (FMQ) buffer, large differences in $f$O$_2$ are manifest as only small changes in the Fe$^{3+}$/$\sum$Fe ratio of the equilibrium silicate liquid \citep{berry_xanes_2003,sossi_redox_2020,hirschmann_magma_2022}, leading to discrepancies in the inferred $f$O$_2$ from a given Fe$^{3+}$/$\sum$Fe ratio. The most prominent example of this disagreement is in the determination of the redox state of Fe in mid-ocean ridge basalt (MORB) glasses. Wet chemical analyses indicate a Fe$^{3+}$/$\Sigma$Fe ratio of 0.10(2) (see \citealt{bezos2021unraveling} and references therein), substantially lower than that determined by Mössbauer spectroscopy; 0.16(1) \citep{cottrell_oxidation_2011}, later revised to 0.143(8) by \citet{zhang_determination_2018}. An alternative Mössbauer-based estimate suggests a considerably lower Fe$^{3+}$/$\Sigma$Fe ratio of 0.10(2) \citep{berry_re-assessment_2018}, in close agreement with wet chemical analyses. \\
%These differing ratios lead to more than an order of magnitude shift in the estimated \textit{f}O$_2$ under which the MORB glasses were thought to have equilibrated \citep{cottrell2021}. 

\noindent In this work, we report the results of \textit{in-situ} high-pressure, room-temperature energy-domain synchrotron Mössbauer spectroscopy (SMS) on peridotitic- and basaltic glasses. The glasses were synthesised under controlled oxygen fugacity to produce exclusively ferrous- and ferric iron-bearing end-members, calibrated independently through colourimetric methods. Doing so enables the unambiguous attribution of the electronic properties of Fe in silicate glasses to their corresponding hyperfine parameters in deduced from Mössbauer spectra. We identify changes in the relative proportions of electronic states of iron as a function of pressure that provide evidence for a high- to low- spin transition in ferrous iron but not in ferric iron. Implications of this result are discussed in the context of defining the redox state of iron in MORB glasses, the terrestrial magma ocean, and rocky planets more massive than the Earth.

%%%%%%%%%%%%%%%%%%%%%%%%%%%%%%%%%%%%%%%%%%%%%%%%%%%%%%%%%%%%%%%%%%%%%%%%%%%%%%%%%%%%%%%%%%%%%%%%%%%%
%%%%%%%%%%%%%%%%%%%%%%%%%%%%%%%%%%%%%%%%%%%%%%%%%%%%%%%%%%%%%%%%%%%%%%%%%%%%%%%%%%%%%%%%%%%%%%%%%%%%
%% Methods

\section{Methods} \label{methods}

%--------------------------------------------------------------------------------
\subsection{Sample preparation}
%--------------------------------------------------------------------------------
The starting materials were synthetic $^{57}$Fe-enriched-peridotitic and -basaltic compositions, prepared from 100$\%$ $^\text{57}$Fe$_2$O$_3$ (95.08$\%$ isotopic enrichment, Isoflex, USA) mixed and ground under acetone in agate mortar from MgO, SiO$_2$, Al$_2$O$_3$ and CaCO$_3$ powders. The powders were based on KLB-1 \citep[PM10-05;][]{davis_composition_2009}, and on a mid-ocean ridge basalt (MORB) composition (BG-01, BG-08). The mixtures were heated in air at 1000°~C, overnight. The $^{57}$Fe-enriched peridotitic glass (PM10-05) was synthesised by placing a fragment of a pellet ($\sim$~20 mg) into a conical nozzle of a laser-heated aerodynamic levitation furnace (see \citealt{badro_experimental_2021}) at the Institut de Physique du Globe de Paris (IPGP), France. An oxygen fugacity of log(\textit{f}O$_2$)~=~-7.99 (PM10-05) was imposed by a flux of CO$_2$-H$_2$, in molar proportions of 25:75, which served to levitate the sample. Heating was achieved by a 125-W continuous wave CO$_2$ laser focused to a spot diameter of $\sim$~2~mm by gold-coated mirrors. The sample was held at 1750~$\pm$~50°~C for 30~sec. Quenching was performed by shutting off the power to the laser, resulting in cooling rates of 800~°C/s, which produced a homogeneous glass without evidence of quench crystallisation, as confirmed by EPMA. \\

\noindent To investigate the differences in the electronic properties of ferrous and ferric iron in end-member compositions under high pressure, fully oxidised and reduced basaltic glasses were synthesised at the Department for Earth and Planetary Sciences (D-EAPS) at ETH Zürich, Switzerland. The reduced basaltic glasses were synthesised in a (1-atm) gas-mixing furnace, at 1400°~C for 4~hours in a gas mixture of 100 ml/min~H$_2$-15.2 ml/min~CO$_2$ metered by Bronkhorst mass flow controllers, resulting in a log(\textit{f}O$_2$)~=~-11.2 \citep[$\Delta$IW~=~-1.5,][]{hirschmann_iron-wustite_2021}. The oxidised basaltic glass was synthesised at 1~GPa (8~\% friction correction applied) and 1400°~C for 4~hours in an end-loaded piston-cylinder in a 3~mm diameter platinum (Pt) capsule, with PtO$_2$ loaded as layers ($\sim$~25 mg) on both the bottom and top of the capsule with basaltic powder ($\sim$~35 mg) sandwiched in between. The PtO$_2$ breaks down to generate an O$_2$ pressure equivalent to that of the total pressure, thereby imposing a log(\textit{f}O$_2$) +~4.8 ($\Delta$IW~+~14.2) given the equation of state of O$_2$ gas \citep{belonoshko_molecular_1991}. A type-B thermocouple (Pt$_{70}$Rh$_{30}$/Pt$_{94}$Rh$_6$) was used to monitor the temperature to an accuracy of $\pm$~5~°C. 

%----------------------------------------------------------------------------------------------
\subsection{Electron-probe microanalysis}

\noindent The major element abundances of the peridotitic and basaltic glasses were determined using a JEOL JXA8230 at the D-EAPS, ETH Zürich. Analyses were performed in wavelength-dispersive mode at 15~kV, 20~nA, with a defocused beam diameter of 10~$\mu$m and using the mean atomic number (MAN) background correction. The calibration was performed using albite (Si), forsterite (Mg), anorthite (Ca, Al), Fayalite (Fe) standards, meanwhile hematite, rutile corundum, periclase and quartz standards were used for the MAN correction. Twenty analyses were collected in a rim-to-rim transverse for the peridotitic glass, whereas a grid of points was used for the basaltic glasses. Each glass was internally homogeneous. The means of the sample compositions and their uncertainties are shown in Table \ref{tab:EPMA_Table}. 

\begin{table}[h]
\centering
  {
    \renewcommand{\arraystretch}{1}
\begin{tabular}{c|c|c|c}
    \hline
        \textbf{Oxide} &  \textbf{PM10-05} & \textbf{BG-08} & \textbf{BG-01}  \\ \hline
        SiO$_{2}$ & 45.7(12) & 54.73(27) & 51.32(18)  \\ 
        MgO & 35.13(12) & 8.04(4) & 7.53(3)  \\ 
        CaO & 4.2(2) & 13.58(6) & 12.5(7)  \\ \
        FeO* & 9.73(7) & 6.21(5) & 10.07(5)  \\ \
        Al$_{2}$O$_{3}$ & 4.77(4) & 16.65(7) & 15.78(6)  \\ 
        PtO & -- & -- & 0.56(2)  \\ 
        SO$_{3}$ & -- & 70 ppm & 120 ppm  \\
        TOTAL & 99.53(18) & 99.22(37) & 97.76(28)  \\ \hline
\end{tabular}
\captionsetup{width=1\textwidth}
\caption[Major element oxide compositions determined by EPMA measurements in peridotitic and basaltic glasses]{Means and standard deviations of major-element oxides determined by EPMA measurements of peridotitic and basaltic glasses (in wt.$\%$ unless otherwise stated). *Iron in Fe$^{3+}$ basaltic glass (BG-01) was recalculated assuming all Fe as Fe$_{2}$O$_{3}$. The lower FeO content in the reduced basaltic glass (BG-08) is attributed to Fe loss during the highly reducing \textit{f}O$_{2}$ synthesis conditions in 1-atm gas mixing furnace.}
\label{tab:EPMA_Table}
}
\end{table} 

%-------------------------------------------------------------------------------------
\subsection{Wet chemical analysis}\label{sec:wetchem}

\noindent The FeO/FeO$_{total}$ ratios of three Fe$^{2+}$-bearing and two Fe$^{3+}$-bearing basaltic glasses were determined using a colourimetric method at the Laboratoire de Planétologie et Géosciences (LPG, Université de Nantes, France) following the methods described in \citet{bezos2021unraveling}. Briefly, roughly 20~mg of the basaltic glasses were ground in an agate mortar under ethanol, of which $\sim$5~mg were digested in 7~mL crystal polypropylene beakers for two days at room temperature in 1~mL distilled HF (40$\%$) and 1~mL of 1.41·10$^{-2}$~molL$^{-1}$ ammonium vanadate solution. The use of ammonium vanadate solution in the "Wilson Method" \citep{wilson1960}, ensures that V$^{5+}$ is in excess of Fe$^{2+}$ during the sample digestion, so as to prevent undesired oxidation of Fe$^{2+}$ in the sample by air. During sample digestion, all Fe$^{2+}$ released from the silicate matrix, is oxidised into Fe$^{3+}$ by the concomitant reduction of V$^{5+}$ to V$^{4+}$ in the low pH acid solution:

\begin{equation}
 V^{\text{5+}} + Fe^{\text{2+}}  \rightleftharpoons V^{\text{4+}} + Fe^{\text{3+}} 
 \label{eq:V-Fe}
\end{equation}

\noindent Following digestion, 5~mL of beryllium sulfate solution was added to the sample solution to remove any excess HF and insoluble fluorides. The solutions were then transferred into volumetric flasks containing 10~ml of ammonium acetate (to increase the pH to $\sim$5 in order to regenerate Fe$^{3+}$, Eq. \ref{eq:V-Fe}), 5~ml of 2:2'‐dipyridil solution and then made up to 100~ml using Milli-Q H$_2$O. Colourimetric measurements of Fe$^{2+}$ and FeO$_{total}$ were made on the same solution by adding 1~ml of hydroxylamine hydrochloride to an aliquant of the FeO solutions, which, being a strong reducing agent, converts all remaining Fe$^{3+}$ into Fe$^{2+}$. We used an ultraviolet/visible spectrophotometer CARY UV500 (Varian) and measured optical densities at 525~nm (where the absorbance by ferrous -2:2’ dipridyl complex is at a maximum) relative to a baseline at 700~nm before and after the addition of the reducing agent \citep{bezos2021unraveling}. The optical density measurements showed that glasses synthesised at $\Delta$IW = -1.5 (BG-08) have an average Fe$^{3+}$/$\Sigma$Fe ratio of 0.02(2), while BG-01 that was synthesised at $\Delta$IW~=~+~14.2 has Fe$^{3+}$/$\Sigma$Fe ratio of 1.00(2). 

%----------------------------------------------------------------------------------------------

\subsection{Synchrotron Mössbauer spectroscopy}
\subsubsection{High pressure experiments}

\noindent The high pressure and room temperature (298~K) experiments were performed using a BX90-type diamond anvil cell \citep{kantor_bx90_2012} equipped with 100~$\mu$m, 150~$\mu$m and 250~$\mu$m-culet diamond anvils. The powdered samples were loaded into a rhenium (Re) gasket without a pressure-transmitting medium. The peridotitic glass was further measured as a single chunk with a Neon (Ne) pressure medium to evaluate the effect, if any, of non-hydrostatic stresses within the cell on hyperfine parameters (Fig. S1 and Table S1). The 250~$\mu$m-thick Re-foil was pre-indented down to 40~$\mu$m for the 250~$\mu$m diamond culet size, 15~$\mu$m for 150~$\mu$m- and 100~$\mu$m diamond culets. A hole of 120~$\mu$m, 60~$\mu$m or 40~$\mu$m, depending on the diamond culet dimensions, was drilled using a laser drill machine that serves as a sample chamber for the powdered glass. 
%Typically, the pressure was increased in steps of 6~-~12~GPa up to 172~-~174~GPa, with estimated uncertainties $\sim$~0.5~-~1~GPa in the low pressure range to $\sim~$3~-~4~GPa in the high pressure range.
The reported pressures, and their uncertainties, were determined before- and after each measurement using the T$_2g$ Raman mode shift of the diamond culet at the center of the compression chamber \citep{akahama_pressure_2006}. \textit{In-situ} high-pressure energy domain SMS measurements of the $^{57}$Fe-enriched peridotitic glasses and basaltic glasses were collected in transmission mode at the nuclear resonance beamline ID18 \citep{potapkin_57fe_2012}, at the European Synchrotron Radiation Facility (ESRF, Grenoble, France), at room temperature. Spectra were collected in the energy-domain with a $\sim$~5$\times$15~$\mu$m beam at an energy of 14.4~keV, during compression and decompression over 1~–~4~h period. The velocity scales were calibrated relative to a natural $\alpha$-Fe foil with 25~$\mu$m thickness at ambient conditions. The linewidth and absolute position of the center shift (CS) relative to $\alpha$-Fe were calibrated against a K$_2$Mg$^{57}$Fe(CN)$_6$ reference single-line absorber, from which the instrumental function \citep{yaroslavtsev_syncmoss_2023} was also extracted.

%--------------------------------------------------------------------------------

\subsubsection{Fitting procedure}

\begin{figure*} [htbp]
      \centering
      \includegraphics[width=0.9\textwidth]{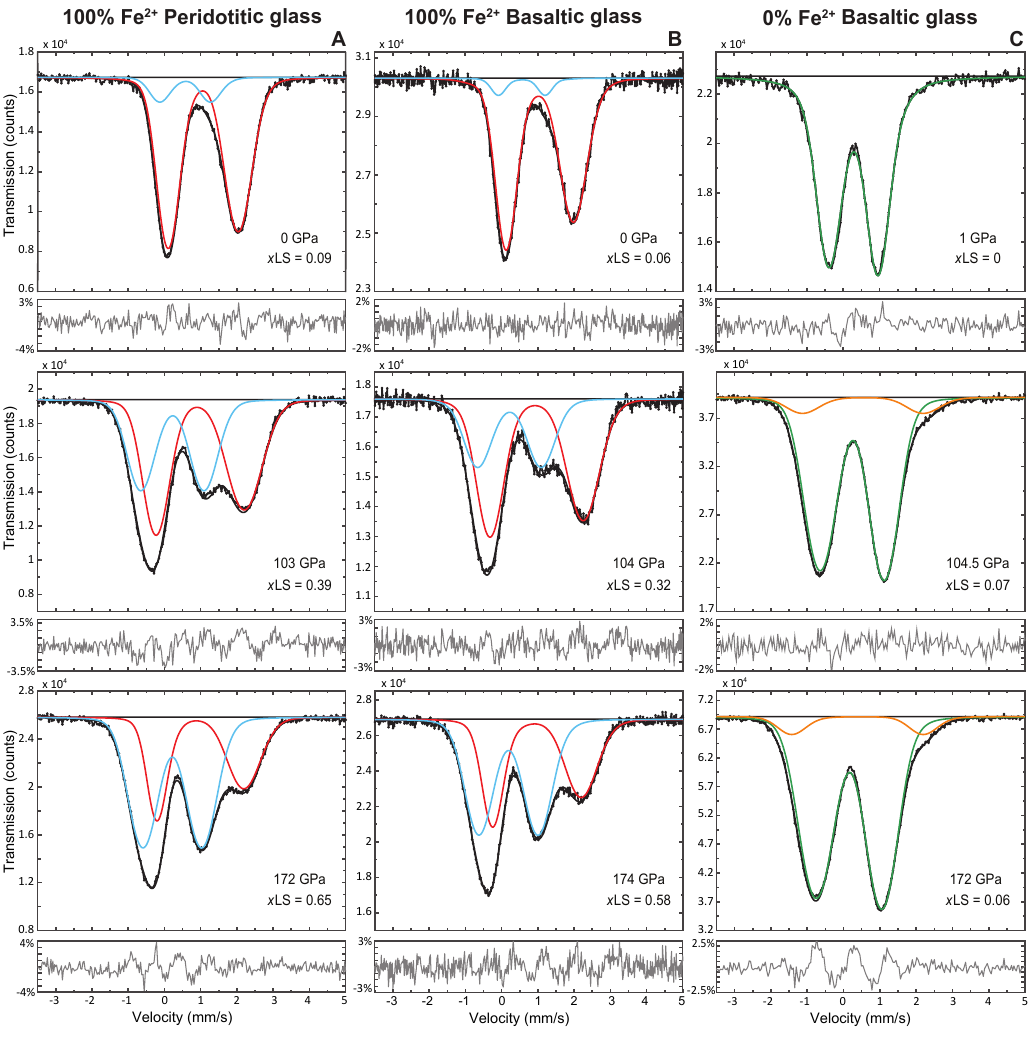}
      \captionsetup{width=1\textwidth}
      \caption[Pressure dependence of synchrotron Mössbauer spectra in peridotitic and basaltic glasses] {Effect of pressure on SMS spectra of $^{57}$Fe-enriched peridotitic and basaltic glasses. (\textbf{A}) Reduced peridotitic glass (PM10-05), (\textbf{B}) reduced basaltic glass (BG-08) and (\textbf{C}) oxidised basaltic glass (BG-01) at room temperature up to 174~GPa. Solid black dots indicate Mössbauer experimental data, and the solid black line shows the fitted curve. (\textbf{A, B}) The red and the light-blue line show Fe$^{2+}$ in the high- and low-spin state (D$_1$ and D$_2$), respectively. (C) The green and the orange line represent the Fe$^{3+}$ in the high- and low-spin, respectively. The fitting residual is shown as a thick grey line beneath each spectrum, expressed in percentage. The pressure at which the Mössbauer spectra were collected and the corresponding low-spin amount (\textit{x}LS) are displayed in the bottom-right corner of each spectrum. (\textbf{A, B}) indicate the \textit{x}LS of Fe$^{2+}$ and (\textbf{C}) \textit{x}LS of Fe$^{3+}$.}
      \label{fig:Spectra_3}
\end{figure*}

\noindent Spectra were fitted using the \texttt{SYNCmoss} software \citep[][Fig. \ref{fig:Spectra_3}, Supplementary Information S1]{yaroslavtsev_syncmoss_2023}.
%, a software package recently developed to correct sample spectra according to the specific response of the instrumental function which characterises synchrotron Mössbauer spectra. 
Mössbauer spectra are fitted by allowing the effective thickness (T), center shift (CS), quadrupole splitting (QS), Gaussian width distribution (G) and the ratio of the Gaussian widths (G$_2$/G$_1$, which if set to unity, yields G = G$_1$) for each doublet to vary (\textit{i.e.,} where 1 and 2 correspond to the peak of each doublet). For the reduced glasses (PM10-05 and BG-08), the D$_1$ doublet was fit over the entire pressure range by allowing T, CS, QS, G and G$_2$/G$_1$ to vary, while Lorentzian width (L) was fixed to 0.098~mm/s (\textit{i.e.,} the natural linewidth) and A (\textit{i.e.,} the ratio between lines due to the texture of the sample) was fixed to 0.5 (\textit{i.e.,} the value that characterises isotropic media). The D$_2$ doublet was fitted using the same approach as for D$_1$, except that G$_2$/G$_1$ which was fixed to 1 over the entire pressure range, because allowing it to vary resulted in values close to 1 in any case. G$_2$/G$_1$ is the ratio of the Gaussian width for two resonance lines, accounting for a linear correlation between CS and QS typical for glasses \citep{virgo_structural_1985, alberto_analysis_1996}, equivalent to the xVBF method \citep{yaroslavtsev2025}. For the oxidised glass, BG-01, spectra collected at ambient pressure up to pressures of 16~GPa were fit by only one doublet, and only beyond 16~GPa was a second doublet introduced, in order to obtain a $\chi^2$ (\textit{i.e.,} goodness-of-fit) close to 1. In all other respects, spectra for BG-01 were fit using the same approach as for PM10-05 and BG-08, excepting that the G$_2$/G$_1$ ratio for the Fe$^{3+}$ doublet was set to 0.9 across the entire pressure range, as linear correlations between CS-QS are typically less pronounced for ferric iron components than they are for ferrous iron \citep{virgo_structural_1985, alberto_analysis_1996}.

%%%%%%%%%%%%%%%%%%%%%%%%%%%%%%%%%%%%%%%%%%%%%%%%%%%%%%%%%%%%%%%%%%%%%%%%%%%%%%%%%%%%%%
\section{Results} \label{results_chapter3}
%----------------------------------------------------------------------------------
\subsection{Reduced glasses} \label{sec:res_reduced_glass_chapter3}

\begin{figure} [htbp]
      \centering
      \includegraphics[width=0.75\textwidth]{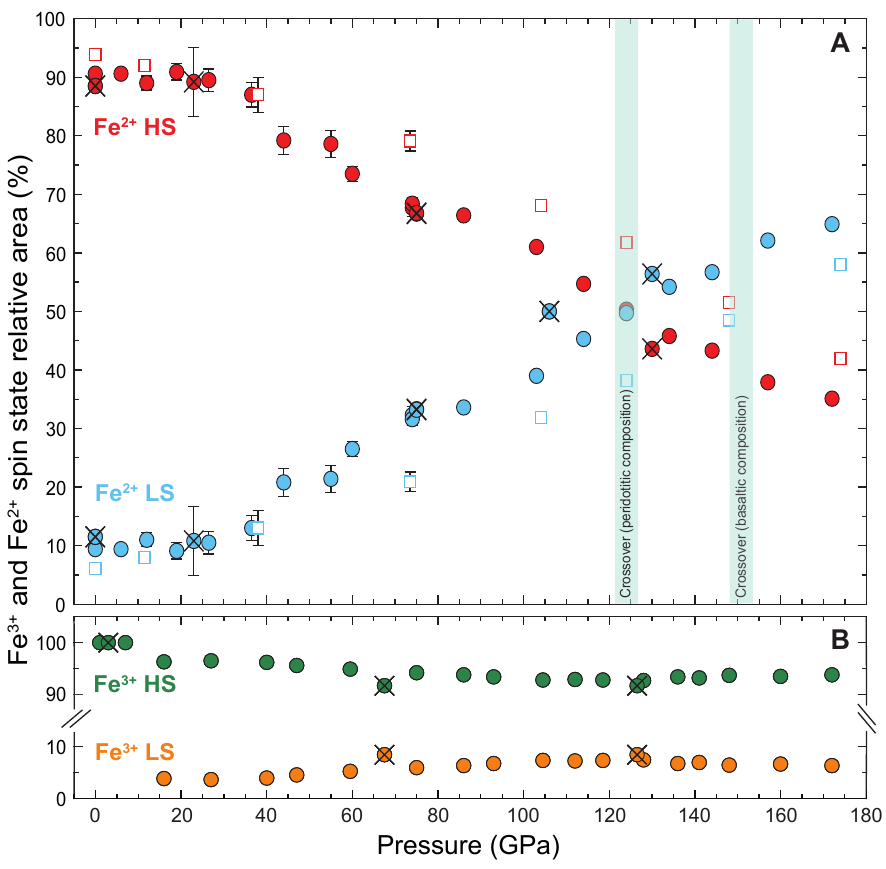}
      \captionsetup{width=1\textwidth}
      \caption[Pressure dependence of the relative areas of Fe$^{2+}$ and Fe$^{3+}$ spin states] {Pressure dependence of the relative areas of Fe$^{2+}$ and Fe$^{3+}$ spin states. (\textbf{A}) Reduced peridotitic glass (PM10-05), circles, and reduced basaltic glass (BG-08), squares. Red denotes Fe$^{2+}$ HS (D$_1$) and light-blue denotes Fe$^{2+}$LS (D$_2$). The two shaded areas refer to the cross-over pressure range for both compositions. (\textbf{B}) Proportions of Fe$^{3+}$ in the high-spin state (green) and in the low-spin state (orange) for oxidised basaltic glass (BG-01). No evidence of HS to LS cross-over in Fe$^{3+}$ has been observed over the pressure range for which the spectra were collected in this study. The circles marked with a cross represent the data collected during decompression. All error bars correspond to one standard deviation on the mean unless otherwise stated. NB: Error bars may be smaller than the size of the points in some cases.}
      \label{fig:Area_All}
\end{figure}

\begin{figure*} [htbp]
      \centering
      \includegraphics [scale=0.9] {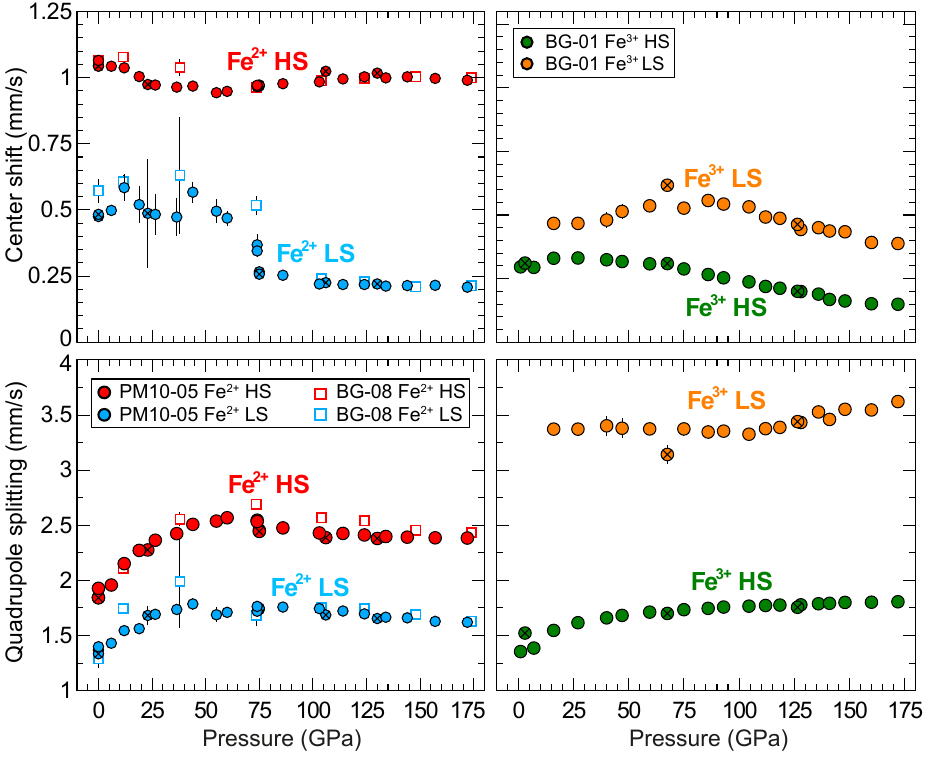}
      \captionsetup{width=1\textwidth}
      \caption[Effect of pressure on the hyperfine parameters of iron in different electronic configurations at high pressure and room temperature] {Effect of pressure on the hyperfine parameters of iron in different electronic configurations at high pressure and room temperature. (Left) Reduced peridotitic (circles) and basaltic glasses (squares) and (Right) oxidised basaltic glass (circles). Data points are colour-coded according to iron oxidation and spin state: Fe$^{2+}$ HS (D$_1$) = red, Fe$^{2+}$ LS (D$_2$) = blue, Fe$^{3+}$ HS = green, Fe$^{3+}$ LS = orange. The center shift (CS) and quadrupole splitting (QS) values obtained from data collected in decompression are indicated by a cross. All error bars correspond to one standard deviation on the mean unless otherwise stated. NB: Error bars may be smaller than the size of the points in some cases.}
      \label{fig:CS_QS_Pressure}
\end{figure*}

\noindent The Mössbauer spectra (Fig. \ref{fig:Spectra_3}) of the two reduced glasses, peridotitic (PM10-05) and basaltic (BG-08), at ambient pressures, were fit with two doublets (D$_1$, red and D$_2$, blue). As pressure increases, the intensity of the D$_2$ doublet increases at the expense of D$_1$, with the relative integrated abundances of the two becoming equivalent at 124~GPa and $\sim$150~GPa (Fig. \ref{fig:Area_All} A) in the reduced peridotitic and basaltic glass, respectively. The two doublets (\textit{i.e.,} electronic configurations) characterise both the reduced glasses throughout the entire pressure range. At 1~bar, the hyperfine parameters of the predominant doublet (D$_1$) ($\sim$~91$\%$ PM10-05; $\sim$~94$\%$ BG-08) yield a center shift (CS) of 1.06~mm/s and quadrupole splitting (QS) of $\sim$1.9~mm/s while the second, weaker doublet (D$_2$) is defined by distinctly lower hyperfine parameter values with a CS~=~0.57~mm/s and QS~=~1.3~-~1.4~mm/s (Fig. \ref{fig:CS_QS_Pressure}). The relative intensity of D$_2$/(D$_1$~+~D$_2$) increases up to 65$\%$ (PM10-05) at 172~GPa and 58$\%$ at 174~GPa (BG-08) at the expense of D$_1$. Furthermore, this change is found to be completely reversible upon decompression (Fig. \ref{fig:Area_All}). The D$_1$ doublet has a CS that maintains a near-constant velocity of $\sim$1~mm/s throughout the pressure, and QS increases up to 2.57~mm/s at 60~GPa in PM10-05, before falling mildly to $\sim$2.4~mm/s by $\sim$170~GPa. For D$_2$, CS ranges from roughly 0.6~-~0.5~mm/s at 0~--~60~GPa, down to 0.21~mm/s by 172~GPa, whereas QS increases to $\sim$1.7~mm/s by 74~GPa before remaining near-constant at this value until $\sim$170~GPa (Fig. \ref{fig:CS_QS_Pressure}, Tables S1-S4). The modest decrease in CS of D$_2$ between 50 -- 75 GPa reflects the large uncertainty in ascertaining its hyperfine parameters at low relative abundances and associated degeneracy in the QS of both D$_1$ and D$_2$. A very similar trend is observed for BG-08. \\

\noindent The hyperfine parameters of the D$_1$ doublet, based on minerals with known stoichiometry and site occupancy, are readily attributable to the high spin (HS) state of Fe$^{2+}$, with our hyperfine parameters being consistent with those determined previously \citep{prescher_iron_2014, murakami_high-pressure_2014, maeda_spin_2017, solomatova_electronic_2017, berry_re-assessment_2018, roskosz_structural_2022}. On the other hand, the attribution of the second doublet in the literature is highly debated, since the hyperfine parameters do not clearly correspond to neither ferrous- nor ferric iron \citep{berry_re-assessment_2018, virgo_structural_1985, jayasuriya_mossbauer_2004,cottrell_high-precision_2009,mashino2025valence}. Because colourimetric measurement of sample BG-08 yields Fe$^{3+}$/$\Sigma$Fe = 0.02~$\pm$~0.02 that is markedly lower than the D$_2$/(D$_1$+D$_2$) ratio of 0.06 (BG-08) or 0.09 (PM10-05), the D$_2$ doublet in these glasses cannot reflect ferric iron, supporting the assertion of \citet{berry_re-assessment_2018}. Importantly, as the bulk composition of the glass does not change during (de)compression at room temperature in our experiments, the attribution of the D$_2$ feature to Fe$^{3+}$ would imply the commensurate oxidation of Fe$^{2+}$ to Fe$^{3+}$ with increasing pressure, and hence the formation of Fe$^0$ (\textit{i.e.,} disproportionation) to conserve oxygen. As there is no evidence of Fe$^0$ in the glass, which would manifest as either a singlet with center shift of -0.3 to -0.5 mm/s \citep[Fe hcp,][]{desmarais2021} or as a sextet with CS $\sim$0 upon quenching (Fe bcc), the D$_2$ feature can be unambiguously assigned to Fe$^{2+}$.\\

\noindent Assignment of the D$_2$ doublet to a discrete coordination environment is less straightforward. The CS and QS values of both Fe$^{2+}$ and Fe$^{3+}$ increase in tandem with coordination number in minerals \citep{mccammon_mossbauer_2021}. Upon compression, the cations in silicate glasses become coordinated by increasingly higher numbers of O atoms due to densification of the glass structure \citep{huang_structural_2022}. Consequently, because D$_2$ has lower CS and QS than does D$_1$, it cannot be attributed to ferrous iron with a higher O coordination number. Instead, its hyperfine parameters suggest a spin transition in Fe$^{2+}$ from  high- to low (LS), as inferred from X-Ray Emission Spectroscopy (XES) of (Mg$_{0.95}$Fe$_{0.05}$)SiO$_3$ glass \citep{nomura2011}. In their case, the transition occurred over $\sim$59--70 GPa, similar to the pressure range over which the fraction of Fe$^{2+}$ LS increases markedly (Fig. \ref{fig:Area_All}). The presence of Fe$^{2+}_{\mathrm{LS}}$ was also observed in magnesiowüstite \citep{speziale_iron_2005}, amorphous olivine \citep{rouquette_high-pressure_2008} and iron magnesium aluminium silicate perovskite \citep{mccammon_iron_2013}. \cite{mashino2025valence} also observe the same spectral evolution in peridotite glass with pressure, but attribute D$_2$ to a combination of Fe$^{2+}$ in intermediate spin (IS) and Fe$^{3+}$ LS. The preceding arguments preclude  Fe$^{3+}$, leaving a spin transition in Fe$^{2+}$ alone as the most likely interpretation. The higher velocity peak of the D$_2$ feature at $\sim$1 mm/s is observed only above $\sim$80~GPa in the reduced glasses (Fe$^{3+}$/$\sum$Fe = 0) of \cite{solomatova_electronic_2017}, their Fig. 5, and is found to grow in intensity above $\sim$55 GPa in the oxidised glasses (Fe$^{3+}$/$\sum$Fe = 0.79) of \cite{mao2014}, their Fig. 1, indicative of qualitative consistency in the D$_2$ feature across Mössbauer spectra of silicate glasses. \cite{mao2014} do not find strong evidence of a spin transition in Fe up to 80~GPa using XES, but this likely results from the high Fe$^{3+}$/$\sum$Fe of their glasses and relatively low pressures of their measurements, such that the total fraction of Fe present as Fe$^{2+}_{\mathrm{LS}}$ in their glass, estimated from our work, is of the order 5~\% and hence not resolvable within their analytical uncertainty.
%A similar transition has been observed in silicate liquids of olivine stoichiometry from $\sim$150~GPa and $\sim$4000~K \citep{shim_ultrafast_2023}. 
Compounds containing Fe$^{2+}_{\mathrm{LS}}$ at ambient pressure are not uncommon, including pyrite, FeS$_2$, with hyperfine parameters not dissimilar to that of D$_2$; QS~=~0.63~mm/s and CS~=~0.3~mm/s \citep{finklea_investigation_1976}. A survey of Fe$^{2+}_{\mathrm{LS}}$-bearing compounds \citep{bancroft1970mossbauer} indicates they exhibit doublets with uniformly low CS (-0.2 to 0.5~mm/s) and QS (0.2 -- 1.8~mm/s), entirely consistent with the range determined for D$_2$ (Fig. \ref{fig:CS_QS_All}). Moreover, Co$^{3+}$, which is isoelectronic with Fe$^{2+}$ (\textit{i.e.,} a 3d$^6$ valence shell), also exists in a non-magnetic low spin state in the octahedral site of the normal spinel \ensuremath{^{IV}\mathrm{Co}^{2+VI}\mathrm{Co}_2^{3+}\mathrm{O}_4} \citep{oneill_thermodynamics_1985,brabers_low-spin-high-spin_1992}. 
%The inferred Fe$^{2+}$-spin transition occurs over a wide pressure range with the pressure cross-over between HS $\rightarrow$ LS at 124~GPa in PM10-05. Hence, at pressures approaching the Earth's present-day core-mantle boundary (136~GPa) at 298~K, ferrous iron is predominantly in its LS state (Fig. \ref{fig:Spectra_3}A). The difference in the cross-over pressure between peridotitic (124~GPa) and basaltic ($\sim$150~GPa) may be related to compositional differences in the glasses (\textit{e.g.,} non-bridging / tetrahedral oxygens and/or FeO content). However, further investigation is needed to confirm and better understand these effects.

%\FloatBarrier
%----------------------------------------------------------------------------------------------

\subsection{Oxidised glass} \label{sec:res_oxidised_glass_chapter3}

\begin{figure} [htbp]
  \centering
  \includegraphics[scale=0.9]{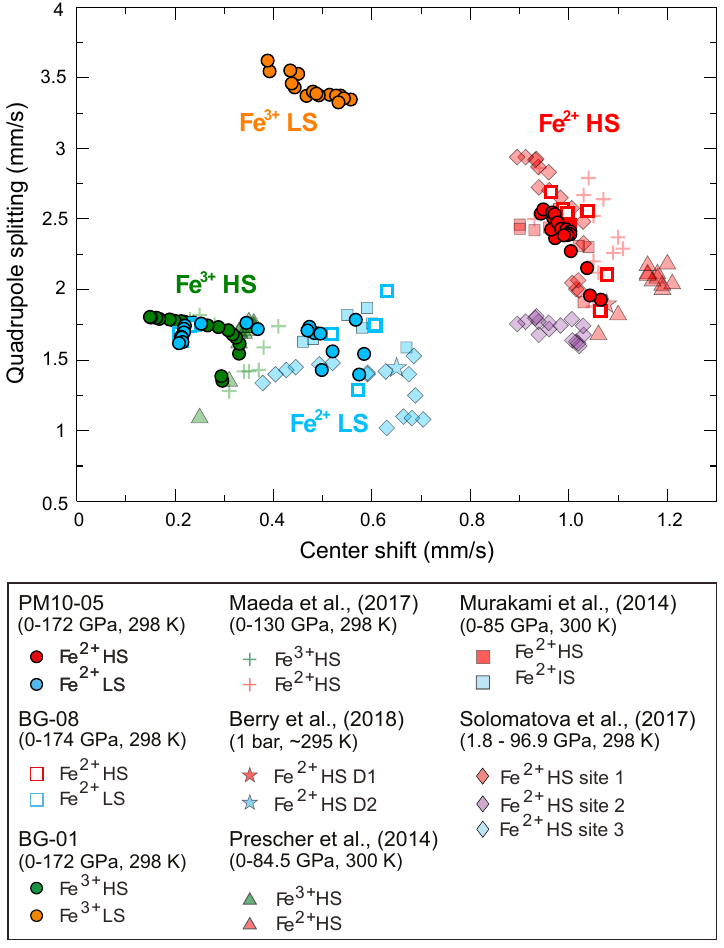}
  \captionsetup{width=1\textwidth}
  \caption [Mössbauer hyperfine parameters of reduced peridotitic- and oxidised basaltic glass at high pressure and room temperature]{Mössbauer hyperfine parameters of reduced peridotitic- and oxidised basaltic glasses at high pressure and room temperature. Full circles and empty-squares = this study, crosses = $^{57}$Fe-enriched basaltic glass \citep{maeda_spin_2017}, stars = $^{57}$Fe- enriched MORB glass \citep{berry_re-assessment_2018}, triangles =  $^{57}$Fe$^ {3+}$ -rich sodium silicate glass without pressure medium \citep{prescher_iron_2014}, full-squares = $^{57}$Fe-enriched enstatite glass \citep{murakami_high-pressure_2014}, diamonds = $^{57}$Fe-basaltic glass \citep{solomatova_electronic_2017}.}
  \label{fig:CS_QS_All}
\end{figure}

\noindent The Mössbauer spectrum of the oxidised glass, BG-01, is adequately fit with one doublet at ambient pressure (Fig. \ref{fig:Spectra_3}C). It has hyperfine parameters (CS~$\sim$~0.29 mm/s, QS~$\sim$~1.35 mm/s at 1~GPa) that are typical for Fe$^{3+}$ \citep[e.g.,][]{mccammon_mossbauer_2021}, meaning iron in BG-01 is entirely ferric, as confirmed by colourimetry (Fe$^{3+}$/$\Sigma$Fe~= 1.00~$\pm$~0.02). With increasing pressure, the CS velocities increase to a peak of $\sim$0.33~mm/s at 27~GPa before decreasing systematically to 0.15~mm/s by 172~GPa, while those for QS increase markedly to $\sim$1.7~mm/s by $\sim$50~GPa, before asymptotically reaching a plateau of $\sim$1.8~mm/s at 172~GPa (Fig. \ref{fig:CS_QS_All}). Such hyperfine parameters resemble those of D$_2$ (Fe$^{2+}$ LS), to the extent that they cannot be reliably separated, particularly at high pressures at which the CS values for Fe$^{3+}$ HS and Fe$^{2+}$ LS converge (Fig. \ref{fig:CS_QS_Pressure}). Indeed, the attribution of a fraction of the D$_2$ feature to Fe$^{3+}$ by \cite{mashino2025valence} in their reduced peridotitic glasses above 60 GPa highlights this point. As such, Fe$^{3+}$/$\Sigma$Fe ratios of glasses with mixed valence cannot be uniquely determined from \textit{in-situ} Mössbauer spectra alone at elevated pressures ($>$~30 GPa). An additional doublet accounts for the growing shoulder at higher velocity beyond 16~GPa, however, its relative intensity never exceeds $\sim$~10$\%$, even at 172~GPa. 
%Although it is not possible to exclude the possibility that this second doublet represents Fe$^{3+}$ HS in a higher coordination environment, its very high quadrupole splitting velocities (3.4~--~3.6~mm/s, Fig. \ref{fig:CS_QS_All}) suggest the presence of Fe$^{3+}$ in the low spin configuration. Should this be the case, a gradual increase in its intensity as a function of pressure would be expected, but is not observed (Fig. \ref{fig:Area_All} B). 
The high QS values of this second doublet (3.4~--~3.6~mm/s, Fig. \ref{fig:CS_QS_All}) depart from those fitted by \cite{sinmyo2017spin} for Fe$^{3+}$ sites in Al-free bridgmanite, but are similar to those inferred for Fe$^{2+}$ in the A-site of bridgmanite \cite{kupenko2015oxidation}, meaning the attribution of this spectral feature remains unclear. 

%%%%%%%%%%%%%%%%%%%%%%%%%%%%%%%%%%%%%%%%%%%%%%%%%%%%%%%%%%%%%%%%%%%%%%%%%%%%%%%%%%%%
\section{Discussion} \label{sec:discussion}
%--------------------------------------------------------------------------------
\subsection{Effect of Fe$^{2+}$ low spin on ambient pressure Mössbauer spectra of silicate glasses}
\label{sec:D2_discussion}

The Fe$^{3+}$/$\Sigma$Fe ratio of mid-ocean ridge basalt (MORB) glasses is used to infer the redox state of their upper mantle sources \citep[e.g.,][]{oneill2018oxidation,cottrell2021}. A variety of measurements converge upon the result that primitive MORB glasses (\textit{i.e.,} those with $>$~8~wt.$\%$ MgO) have roughly 10$\%$ Fe$^{3+}$/$\Sigma$Fe 
\citep{oneill2018oxidation,bezos2021unraveling,oneill2024}. Indeed, Mössbauer spectra of basaltic glasses synthesised near the FMQ buffer bear similarities to those of MORB glasses \citep{cottrell_high-precision_2009,berry_re-assessment_2018}. In both studies, the D$_1$ and D$_2$ features are resolvable, and, as deduced herein, the latter amounts to 7.5~$\pm$~1.5~\% of the total iron in the glass. While \citet{cottrell_high-precision_2009} ascribe the D$_2$ feature to Fe$^{3+}$, \citet{berry_re-assessment_2018} propose that it reflects Fe$^{2+}$, leading to discrepant estimates of Fe$^{3+}$/$\Sigma$Fe in MORB of 0.16(1) and 0.10(2), respectively. \citet{zhang_determination_2018} indicated that some discrepancy arises from differences in the recoil-free fraction of Fe$^{2+}$ and Fe$^{3+}$, leading to a slight overestimation of the Fe$^{3+}$/$\Sigma$Fe ratio at room temperature from 0.16(1) to 0.143(8). \\
%Nevertheless, a difference of roughly 0.04 still remains between the two interpretations of the Mössbauer data.

\noindent Here we show that the D$_2$ feature can be unambiguously assigned to Fe$^{2+}$, likely in its low-spin state, supporting the contention that the Fe$^{3+}$/$\Sigma$Fe of MORB glasses is 0.10(2) \citep{berry_re-assessment_2018,bezos2021unraveling}. As such, calibrations ascribing D$_2$ to Fe$^{3+}$ for Mössbauer spectra measured at ambient pressure and temperature (1 bar, 298 K) require a further downwards correction on the apparent (measured) ferric iron fraction, Fe$^{3+}/\Sigma$Fe$_{meas}$:

\begin{equation}
    1-\left(\frac{\mathrm{Fe}^{3+}}{\Sigma \mathrm{Fe}}\right)_{corr} = \frac{1 - \left(\frac{\mathrm{Fe}^{3+}}{\Sigma \mathrm{Fe}}\right)_{meas}}{R}
    \label{eq:D2_corr}
\end{equation}

\noindent where $corr$ denotes the corrected Fe$^{3+}/\Sigma$Fe of the glass and $R = \frac{D_1}{[D_1~+~D_2]}_{red}$ is the integrated area ratio of D$_1$ to the sum of D$_1$ and D$_2$ for entirely reduced glasses from this work, 0.94(1) for basaltic glass and 0.91(1) for peridotitic glass. This correction assumes that the ratio $\frac{D_1}{[D_1~+~D_2]}$ is independent of the Fe$^{3+}$/$\Sigma$Fe ratio of the glass, and states that the corrected Fe$^{2+}$/$\Sigma$Fe of the glass is equal to that measured divided by $R$. The correction has limits of Fe$^{3+}/\Sigma$Fe$_{corr}$~=~Fe$^{3+}/\Sigma$Fe$_{meas}$ when Fe$^{3+}/\Sigma$Fe$_{meas}$~=~1, and Fe$^{3+}/\Sigma$Fe$_{corr}$~=~0 as Fe$^{3+}/\Sigma$Fe$_{meas}$ approaches $\frac{D_1}{[D_1~+~D_2]}_{red}$. \\

\citet{zhang_ferric_2024} collected energy-domain Mössbauer spectra on quenched glasses synthesised at pressures ranging from 38 to 71~GPa and 3600 to 4400~K. The nearly constant hyperfine parameters observed across all spectra reported by \citet{zhang_ferric_2024} are consistent with those for Fe$^{2+}$ and Fe$^{3+}$ species at 1~bar and 298~K; the conditions of their measurements. This observation indicates that Fe$^{2+}$ LS observed in our glasses is not quenchable to ambient conditions, as confirmed by the reversibility of the Mössbauer spectra with pressure (Fig. \ref{fig:Area_All}). Because the D$_2$ feature is present, even at 1~bar and 298~K, Eq. \ref{eq:D2_corr} is used to correct the Fe$^{3+}$/$\Sigma$Fe of the glasses reported in \citet{zhang_ferric_2024}. To do so, we use the $\frac{D_1}{[D_1~+~D_2]}$ determined for Fe$^{2+}$ basaltic glass (0.94), and is therefore a minimum estimate. At the modest Fe$^{3+}/\Sigma$Fe$_{meas}$ reported in \citet{zhang_ferric_2024}, 0.204 to 0.440, the corrections are between 0.034 and 0.048 absolute (see Table S6). 

%----------------------------------------------------------------------------------------------

\subsection{\texorpdfstring{Gibbs energy change of the high- to low- spin transition in Fe$^{2+}$-bearing glasses}{Gibbs energy change of the high- to low- spin transition in Fe2+-bearing glasses}}
\label{sec:g_spin}

\noindent If, as suggested, the electronic environment of D$_2$ is Fe$^{2+}_{\mathrm{LS}}$, then the transition from D$_1$ to D$_2$ can be described by the reaction:

\begin{equation}
    \mathrm{FeO_{HS}} = \mathrm{FeO_{LS}}
    \label{eq:Fe2-LS-HS}
\end{equation}

\noindent because the HS and LS states are the only two present in the glass, the corresponding Gibbs free energy change is written:

\begin{equation}
    \Delta G_{(\ref{eq:Fe2-LS-HS})} = -RT\mathrm{ln}\left(\frac{x\mathrm{FeO_{LS}}}{[1-x]\mathrm{FeO_{LS}}}\right).
    \label{eq:Fe2-LS-HS-dG}
\end{equation}

\noindent An expression describing $ \Delta G_{(\ref{eq:Fe2-LS-HS})}$ is required. 
%Although temperature is constant in the experiment, as iron is a major element in the glasses, it cannot be assumed \textit{a priori} that the thermodynamic potentials ($H$, $S$ and $V$) of Fe$^{2+}$ LS and  Fe$^{2+}$ HS are independent of pressure (cf. changes in their hyperfine parameters, Fig. \ref{fig:CS_QS_Pressure}). Such changes may arise because Fe$^{2+}$, as other M$^{2+}$ cations, occupies a distribution of sites in the silicate glass (\textit{e.g.,} \citealt{wilke2005,lelosqsossi2023}), each coordinated by O$^{2-}$ anions, themselves coordinated to silicate- and aluminate units (typically SiO$_4^{4-}$). In order to explore the consequences of spin transitions in disordered phases, we devise a model in which the 
Here, it is expressed as a polynomial function of the 1 bar Gibbs energy and volume at constant $T$. Relative to a standard state of $x = 0$ (\textit{i.e.,} no LS), $G$ is composed of two terms;
\begin{enumerate}
    \item A configurational entropy contribution, supposing that the LS state simply replaces the HS state on defined sites;

    \begin{equation}
        \Delta S^{ideal}_{config} = -RN\left(x\mathrm{ln}(Nx) + (1-x)\mathrm{ln}(N[1-x])\right)
        \label{eq:S_config}
    \end{equation}

    where $N$ is the mole fraction of FeO which describes ideal (random) mixing on octahedral Fe$^{2+}$ sites capable of undergoing a spin transition, and

    \item A non-configurational contribution to $G$,

    \begin{equation}
        \begin{split}
        G_{non\text{-}config}(T, P, x) = Nx\Big( 
        \Delta H_{LS-HS}[T, 1~\mathrm{bar}, x] \\
        - T \Delta S_{non\text{-}config}[T, 1~\mathrm{bar}, x]
        + \int_1^P \Delta V_{LS-HS} dP [T, P, x]\ \Big)
    \end{split}
    \label{eq:G_non-config}
    \end{equation}

    and $G = G_{non-config} - T \Delta S^{ideal}_{config}$.\\
    
\end{enumerate}

\noindent The zeroth-order approximation is to assume that $\Delta H$ (1 bar, $x$), $\Delta S$ (1 bar, $x$) and $\Delta V$ ($P$, $x$) are independent of temperature. Secondly, because all measurements were performed at constant temperature (300~K), $\Delta H$ and $\Delta S$ cannot be uniquely determined. Finally, we assume $G$ and $\Delta V$ are only functions of $x$, since $P$ and $x$ are highly correlated (Fig. \ref{fig:Area_All}). Therefore, the simplified expression for $G$ is given by polynomial expansion, typical for Landau theory-type expressions for free energy \citep{carpentersaljie1994}:

\begin{equation}
    G_{non-config} = Nx (g_0 + g_1x + g_2x^2 + g_3x^3 + ...) + NPx(v_0 + v_1x + v_2x^2 + v_3x^3 + ...)
    \label{eq:G_polynomial}
\end{equation}

\begin{figure}[!ht]
        \centering
        \includegraphics[width=1\linewidth]{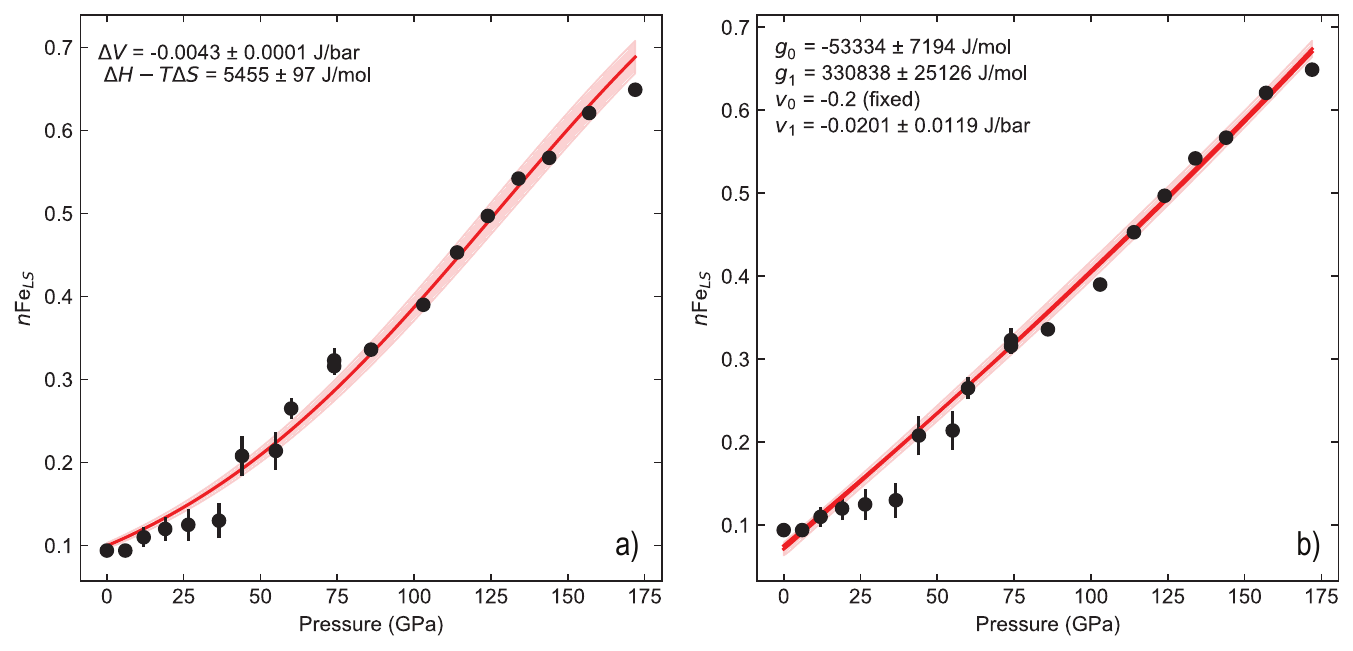}
        \captionsetup{width=1\textwidth}
        \caption{The fraction of ferrous iron in its low spin state ($n$Fe$_{LS}$) as a function of pressure (in GPa) as determined for the reduced peridotitic glass, PM10-05 at 298~K (black points). The red curve is a error-weighted least-squares fit to the data using a) Eq. \ref{eq:Fe2-LS-HS_dV} and b) Eq. \ref{eq:G_equilib} and their 1s error envelope (red field). The best-fit values of $\Delta$V and $\Delta$H $-~T \Delta$S (a) and $g_0$, $g_1$, $v_0$ (fixed) and $v_1$ (b) are also shown.}
        \label{fig:nFe_LS}
\end{figure}

\noindent The condition for equilibrium at constant pressure ($P$), temperature ($T$) and mole fraction of FeO ($N$) is:

    \begin{equation}
        \begin{split}
        \left( \frac{ \partial G } {\partial x} \right)_{T, P, N} &= 0 =  RT\ln\left(\frac{x\,\mathrm{FeO_{LS}}}{[1-x]\,\mathrm{FeO_{LS}}}\right) + g_0 + 2g_1x + 3g_2x^2 + \dots \\
         &\quad + P\left(v_0 + 2v_1x + 3v_2x^2 + \dots\right)
        \end{split}
        \label{eq:G_equilib}
    \end{equation}

\noindent where $g_n$ and $v_n$ are the $n^{th}$-order polynomial expansion coefficients of the molar Gibbs Free Energy and volume of reaction \ref{eq:Fe2-LS-HS} at 300~K. Note that the mole fraction ($N$) cancels. Truncating Eq. \ref{eq:G_equilib} to the first terms of the expansion (\textit{i.e.,} $g_0$ and $v_0$) yields:

\begin{equation}
    \Delta G_{(\ref{eq:Fe2-LS-HS})} = \int^P_{1} \Delta V_{LS-HS} \mathrm{\textit{d}}P + C
    \label{eq:Fe2-LS-HS_dV}
\end{equation}

\noindent where the integration constant, $C = \Delta G_{non-config} = \Delta H - T \Delta S$, and implies the volume change ($\Delta$V) and the 1 bar non-configurational Gibbs energy change ($\Delta G_{non-config}$) are independent of pressure. Values of $\Delta V_{LS-HS}$ = -0.0043~$\pm$~0.0001~J/bar and  $\Delta H - T \Delta S$ = 5455~$\pm$~97~J/mol are found to fit the data satisfactorily within uncertainty (Fig. \ref{fig:nFe_LS}a). \\

\noindent Spin transitions identified for crystalline phases have $\Delta V_{LS-HS}$ of the order of -0.08 J/bar (ferropericlase at 300~K; \citealt{komabayashi_high-temperature_2010}) or -0.2~J/bar (Co$_3$O$_4$ spinel at 1 bar; \citealt{oneill_thermodynamics_1985}). Similar estimates are found from tabulated ionic radii of $^{VI}$Fe$^{2+}$ HS and LS (0.78~{\AA} and 0.61~{\AA}, respectively) for a hypothetical spin transition in wüstite ($r_\mathrm{O}$ = 1.40~{\AA}, rock salt structure). This yields a molar volume of $V^{FeO}_{HS}$~=~1.25~J/bar close, but not identical to the 1.20~J/bar value reported in \citet{robiehemingway1995}, due to non-stoichiometry, and $V^{FeO}_{LS}$~=~0.98~J/bar, thus, $\Delta$V$^{FeO}_{LS-HS}$ is likely -0.2 to -0.3~J/bar; much more negative than that found from fits to the data using Eq. \ref{eq:Fe2-LS-HS_dV}. This leaves two possibilities; either \textit{i)} the electronic transition observed in the glasses is not associated with a transition from high- to low spin or \textit{ii)} the assumption of constant $\Delta$V and $\Delta G_{non-config}$ with pressure is invalid.\\

%However, because this solution returns an unphysical volume, and the distribution of sites in the glass implies $g$ is a function of $P$ and $x$, 
\noindent To test option \textit{(ii)}, we include the next terms in the polynomial expansion; $g_1$ and $v_1$. Such terms are to be expected, because Fe$^{2+}$ atoms in their HS state at ambient pressure occupy a distribution of coordination environments in the glass (\textit{e.g.,} \citealt{wilke2005, lelosqsossi2023}) such that the energy required to change Fe$^{2+}$ HS to Fe$^{2+}$ LS will depend on their local environment. 
%For example, an octahedrally coordinated Fe$^{2+}$ HS atom will require less energy to transition to the LS state than would a tetrahedrally coordinated one.
%atoms will be drawn into octahedral coordination at different pressures, and 
This coordination environment should also become more distorted with increasing pressure, as indicated by the increasing QS values of the fitted doublets (Fig. \ref{fig:CS_QS_Pressure}). Therefore the data are now fit accounting for the next-simplest description of $g$, that is, including $g_0$, $g_1$, $v_0$ and $v_1$, with the additional constraint that $v_0$~=~-0.2~J/bar.\\

\noindent The best fit values are $g_0 = -53334 \pm 7194$~J/mol, $g_1 = 330838 \pm 25126$ J/mol, $v_0 = -0.2$ J/bar (fixed) and $v_1 = -0.020 \pm 0.012$~J/bar (Fig. \ref{fig:nFe_LS}b). This illustrates that $v_0$~=~-0.2 J/bar, as expected for an HS $\rightarrow$ LS transition, is compatible with the interpreted increase in LS iron in the silicate glasses with pressure, provided $v$ and $g$ are allowed to vary with $x$Fe$^{2+}_{\mathrm{LS}}$. The volume change of Eq. \ref{eq:Fe2-LS-HS} in this scheme varies modestly with $x$, where $v_0 + 2v_1x$ = -0.2 J/bar at $x = 0$ to -0.227 J/bar at $x$~=~0.675. Over this range of $x$, $g_0 + 2g_1x$ varies from -6500~J/mol at $x = 0$ to 392000~J/mol at \textit{x} = 0.675, indicating the larger (\textit{i.e.,} more negative) $\Delta V_{\mathrm{LS-HS}}$ compared to that derived from Eq. \ref{eq:Fe2-LS-HS_dV} is compensated by a large increase in $g_0 + 2g_1x$ with $x$.\\

\noindent Part of the Gibbs energy change arises from the electronic entropy of the spin-pairing and the change in molar volume. By analogy with the LS-to-HS transition of isoelectronic Co$^{3+}$ in Co$_3$O$_4$ spinel \cite{oneill_thermodynamics_1985}, the entropy change would be ~$\sim$20 J/K per mole of Fe$^{2+}$ (see below). At 300~K, this equates to 6000 J/(mol.K) and is hence $\sim$60~$\times$ smaller than that required by the value of $g_0 + 2g_1x$ at high $x$ and 300~K. Hence, the large value of $g_1$ reflects an increase in enthalpy as a function of $x$ and $P$, which could reflect gross distortions in the geometries of individual cation sites (e.g., \citealt{gilletrichet1997}). Because the mole fraction of FeO in the glasses is $\sim$0.1, the equivalent enthalpy change ($\Delta H_{\mathrm{LS-HS}})$ per mole of glass is correspondingly ten times lower, $\sim$30~kJ/mol. This is reasonable to first order, given the $\Delta H_{298}$ of formation from the oxides of common silicate melt components; for example, 0.5 MgO + 0.5 SiO$_2$ $\Leftrightarrow$ 0.5 MgSiO$_3$ is -19.65~kJ/mol \citep{robiehemingway1995}.

%%%%%%%%%%%%%%%%%%%%%%%%%%%%%%%%%%%%%%%%%%%%%%%%%%%%%%%%%%%%%%%%%%%%%%%%%%%%%%%%%%%%%
\subsection{A goldilocks pressure for Earth and suppression of oxidation in super-Earths}

The inferred stability of Fe$^{2+}_{\mathrm{LS}}$ in silicate glasses suggests it may also occur in their molten counterparts. Indeed, \textit{ab-initio} simulations show that ferrous iron is entirely in its low-spin state at 140 GPa at 4000~K in pyrolitic liquid \citep{caracas_meltcrystal_2019}, a prediction borne out by \textit{in-situ} X-Ray Emission spectra of olivine liquids at 150~GPa \citep{shim_ultrafast_2023}. Yet, the effect of this transition on the redox state of the Earth's mantle, and those of other planetary bodies remains unconstrained. Here, we devise a model to predict the effect of the spin transition in ferrous iron on the resulting Fe$^{3+}/\Sigma$Fe of peridotitic liquid at 4000~K. This ratio is related to $f$O$_2$ by the equilibrium:

\begin{equation}
    \mathrm{FeO} + \frac{1}{4}\mathrm{O}_2 = \mathrm{FeO}_{1.5}.
    \label{eq:Fe2-Fe3-fO2}
\end{equation}

\begin{figure}[!ht]
    \centering
    \includegraphics[width=0.5\linewidth]{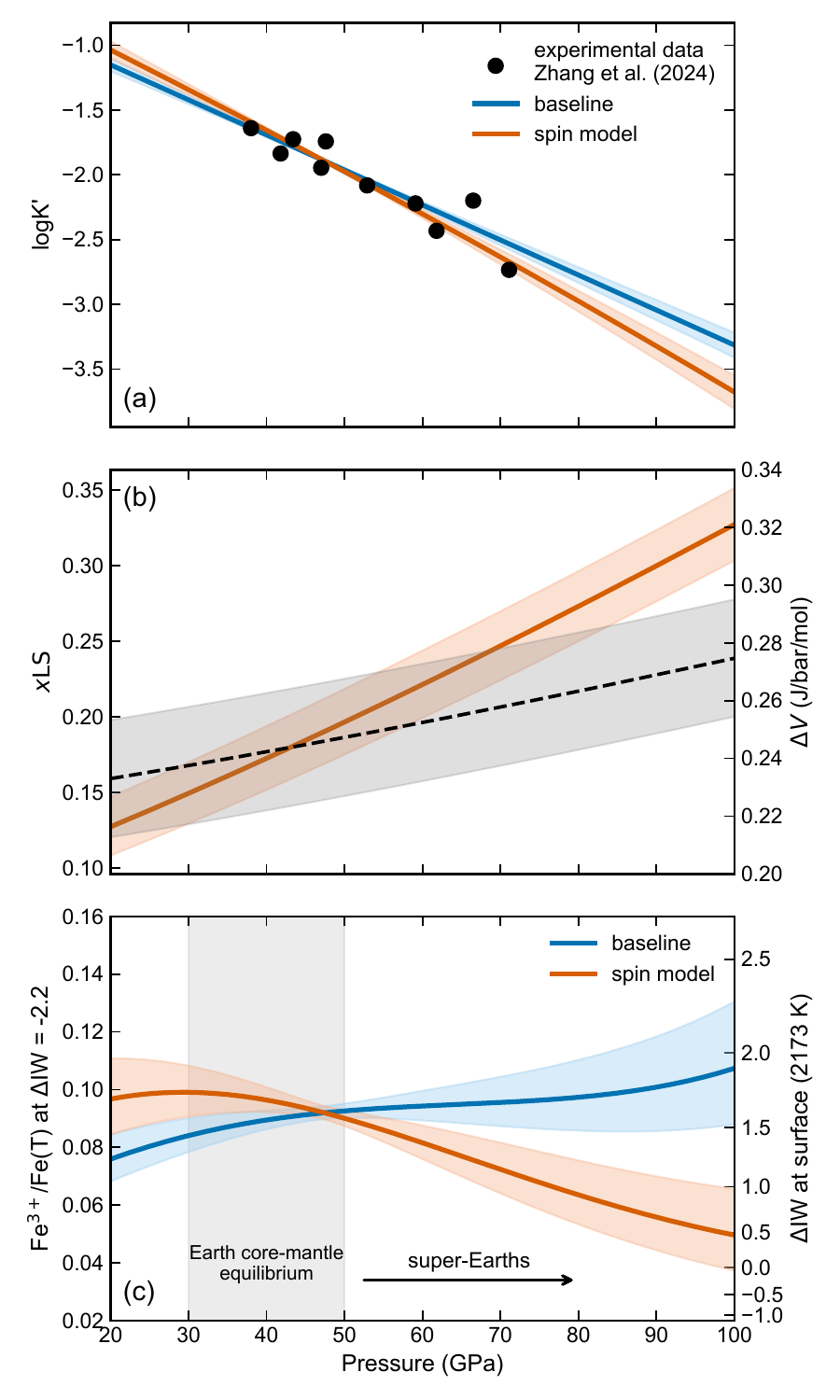}
    \caption{Properties of the reaction FeO + $\frac{1}{4}$O$_2$ = FeO$_{1.5}$ in deep magma oceans at 4000~K. (a) Model fits to the equilibrium constant, $\mathrm{log}K'_{(\ref{eq:Fe2-Fe3-fO2})}$, as a function of pressure from the experiments of \cite{zhang_ferric_2024}, using eq. \ref{eq:Fe2-Fe3-fO2_HS_fit} (baseline - blue line) and eq. \ref{eq:spin_fit} (spin model - orange curve). (b) The proportion of Fe$^{2+}$ in the low spin (LS) state for the spin model (orange solid line) and the net volume change of the reaction (black dashed line), $\Delta$V$_{(\ref{eq:Fe2-Fe3-fO2})}$ = $\Delta$V$_{(\ref{eq:Fe2-Fe3-fO2_HS})}$ - $x$LS$\times$$\Delta$V$_{(\ref{eq:Fe2-LS-HS})}$. (c) The computed Fe$^{3+}$/$\sum$Fe ratio of the silicate liquid at 4000~K for a fixed $f$O$_2$ of -2.2 log$_{10}$ units relative to the iron-wüstite (IW) buffer \citep{hirschmann_iron-wustite_2021}, typical of core formation on Earth \citep{frost_redox_2008}. The secondary $y$-axis is computed by converting the Fe$^{3+}$/$\sum$Fe ratio into an $f$O$_2$ at 1 bar and 2173 K using the calibration of \cite{sossi_redox_2020}. The grey field denotes the mean pressure of core-mantle equilibration. The envelopes denote the 1-$\sigma$ uncertainties that include the uncertainty on the fit (baseline) as well as that on $\Delta S_{\mathrm{non-config}}$ (spin model).}
    \label{fig:spin_model}
\end{figure}

\noindent In order to compute the apparent equilibrium constant, $K'_{(\ref{eq:Fe2-Fe3-fO2})}$, of this reaction, we use the experiments of \cite{zhang_ferric_2024}, for which Fe$^{3+}/\Sigma$Fe$_{meas}$ (corrected according to our Mössbauer doublet assignment, see section \ref{sec:D2_discussion}) of the glasses were determined and the $f$O$_2$ under which the experiment equilibrated is constrained:

\begin{equation}
    \mathrm{log}K'_{(\ref{eq:Fe2-Fe3-fO2})} = \mathrm{log}\left[\frac{x\mathrm{FeO}_{1.5}}{x\mathrm{FeO}.(f\mathrm{O}_2)^{0.25}}\right] = \frac{-\Delta G_{(\ref{eq:Fe2-Fe3-fO2})}}{2.303 RT},
    \label{eq:Eq_Constant}
\end{equation}

\noindent where $x$ is the mole fraction. This value ranges between -2.73 at 71 GPa and -1.64 at 38 GPa, at an average temperature of 4070$\pm$170~K \citep{zhang_ferric_2024} and is negatively correlated with pressure (black points, Fig. \ref{fig:spin_model}a). However, this expression does not specify the spin state of FeO or FeO$_{1.5}$. To explicitly account for the effect of the spin transition on $\mathrm{log}K'_{(\ref{eq:Fe2-Fe3-fO2})}$, constraints on eq. \ref{eq:Fe2-LS-HS} at  300~K must be extrapolated to 4000~K. The low-spin state is the low-entropy (and hence low-temperature) electronic configuration due to i) the pairing of electrons in the lower-energy orbitals of the crystal field split (roughly -9~J/[K.mol], see section \ref{sec:g_spin}) and ii) its smaller ionic radius compared to the high-spin state \citep{oneill_thermodynamics_1985}:

\begin{equation}
    \partial S = \gamma C_v \frac{\partial V}{V},
    \label{eq:S_v_collapse}
\end{equation}

\noindent where $\gamma$ is the Grüneisen parameter (taken to be 1.5, typical for silicates \citealt{asimow2012shock}), $C_v$ is the heat capacity at constant volume (roughly $3nR$ for solids) and $n$ the number of atoms (=2). This yields a contribution of $\partial S \simeq 1.5 \times 6 \times 8.314 \times -0.2/1.25 = -12$ J/(mol.K), and hence a total non-configurational entropy change, $\Delta S_{non-config}$ of roughly -20 J/(mol.K). This permits a complete description of eq. \ref{eq:G_equilib} as:

\begin{equation}
        \left( \frac{ \partial G } {\partial x} \right)_{T, P, N} = 0 = RT\ln\left(\frac{x\,\mathrm{FeO^{LS}}}{[1-x]\,\mathrm{FeO^{LS}}}\right) + g_0 + Pv_0 - (T - T_{\mathrm{ref}})\Delta S_{non-config} + 2[g_1 + Pv_1]x,
        \label{eq:G_equilib_dS}
\end{equation}

\noindent where $T_{ref}$ = 300~K. The corresponding Gibbs energy change of eq. \ref{eq:Fe2-LS-HS} is:

\begin{equation}
        \Delta G_{ (\ref{eq:Fe2-LS-HS})} = x[g_0 + Pv_0 - (T - T_{\mathrm{ref}})\Delta S_{non-config}] + [g_1 + Pv_1]x^2 + RT[x \mathrm{ln}x + (1-x) \mathrm{ln}(1-x)].
        \label{eq:G_equilib_dS_dG}
\end{equation}

Eq. \ref{eq:Fe2-Fe3-fO2} can be written explicitly for the HS state:

\begin{equation}
    \mathrm{FeO^{HS}}+ \frac{1}{4}\mathrm{O}_2 = \mathrm{FeO^{HS}_{1.5}}.
    \label{eq:Fe2-Fe3-fO2_HS}
\end{equation}

\noindent Recognising that the data is consistent with a linear fit in $\mathrm{log}K'_{(\ref{eq:Fe2-Fe3-fO2})}$ vs. $P$, we make the assumption that eq. \ref{eq:Fe2-Fe3-fO2_HS} has a constant $\Delta V_{\mathrm{FeO^{HS}_{1.5}-FeO^{HS}}}$ over the pressure interval where Fe$^{2+}_{\mathrm{LS}}$ should be minor ($<$60 GPa):

\begin{equation}
    \Delta G_{ (\ref{eq:Fe2-Fe3-fO2_HS})}(P, 4000~\mathrm{K}) = \Delta G_{ (\ref{eq:Fe2-Fe3-fO2_HS})}(1~\mathrm{bar}, 4000~\mathrm{K})  + \Delta V_{\mathrm{FeO^{HS}_{1.5}-FeO^{HS}}} P,
    \label{eq:Fe2-Fe3-fO2_HS_fit}
\end{equation}

where $\Delta G_{ (\ref{eq:Fe2-Fe3-fO2_HS})}(1~\mathrm{bar}, 4000~\mathrm{K})$ and $\Delta V_{\mathrm{FeO_{1.5}-FeO}}$ are given by fits to the data, yielding 46710$\pm$19340 J/mol and 0.207$\pm$0.041 J/bar/mol, respectively. Over the same pressure range, this volume change overlaps with that deduced by \cite{deng_magma_2020}, $\sim$0.19 J/bar/mol (their Fig. 1a) and by \cite{zhang_ferric_2024}, $\sim$0.22 J/bar/mol (their Fig. S5). However, the transition of high- to low-spin ferrous iron (eq. \ref{eq:Fe2-LS-HS}) implies $\Delta V_{\mathrm{FeO_{1.5}-FeO}}$ changes with pressure. Hence, by subtracting eq. \ref{eq:Fe2-LS-HS} from eq. \ref{eq:Fe2-Fe3-fO2_HS}, an expression for $\mathrm{log}K'_{(\ref{eq:Fe2-Fe3-fO2})}$ that accounts for the effect of spin transition in FeO is given:

\begin{equation}
    \mathrm{log}K'_{(\ref{eq:Fe2-Fe3-fO2})}(P, 4000~\mathrm{K}) = \frac{-\Delta G_{(\ref{eq:Fe2-Fe3-fO2})}}{2.303 RT} = -\left(\frac{\Delta G_{ (\ref{eq:Fe2-Fe3-fO2_HS})} - \Delta G_{(\ref{eq:Fe2-LS-HS})}}{2.303 RT}\right).
    \label{eq:spin_fit}
\end{equation}

\noindent Fits to the data using eq. \ref{eq:Fe2-Fe3-fO2_HS_fit} (`baseline') and eq. \ref{eq:spin_fit} (`spin model') are shown in Figs. \ref{fig:spin_model}a-c, with their 1-$\sigma$ envelopes that account for the uncertainties due to the scatter in the data of \cite{zhang_ferric_2024} and in $\Delta S_{non-config}$, taken to be $\pm$ 5 J/(mol.K).

Two key observations result. The first is that the inferred pressure for core-mantle equilibrium on Earth $\sim$30--50 GPa \citep[e.g.,][]{siebert2012} coincides with the peak of the Fe$^{3+}$/$\sum$Fe (10~\%) for a given $f$O$_2$ ($\Delta$IW = -2.2). The corollary is that the Earth may have a Goldilocks mass that resulted in mean pressures of core-mantle equilibrium suitable for maximising the Fe$^{3+}$/$\sum$Fe of the mantle near $\sim$10~\%. Second, super-Earths, should have lower Fe$^{3+}$, by at least a factor of two, than Earth's mantle for the same $f$O$_2$ relative to IW (5~\% at $\Delta$IW = -2.2 and 100 GPa). Assuming the Fe$^{3+}$/$\sum$Fe set by core-mantle equilibrium is constant throughout the extent of the mantle, the calibration of \cite{sossi_redox_2020} illustrates that this difference implies an $f$O$_2$ at the surface that is nearly 2 log$_{10}$ units lower (Fig. \ref{fig:spin_model}c). Consequently, oxidation of super-Earth- and sub-Neptune mantles should be inhibited.

\subsection{Implications for atmospheres on Earth-like planets, super-Earths and sub-Neptunes}

\begin{figure}
    \centering
    \includegraphics[width=0.75\linewidth]{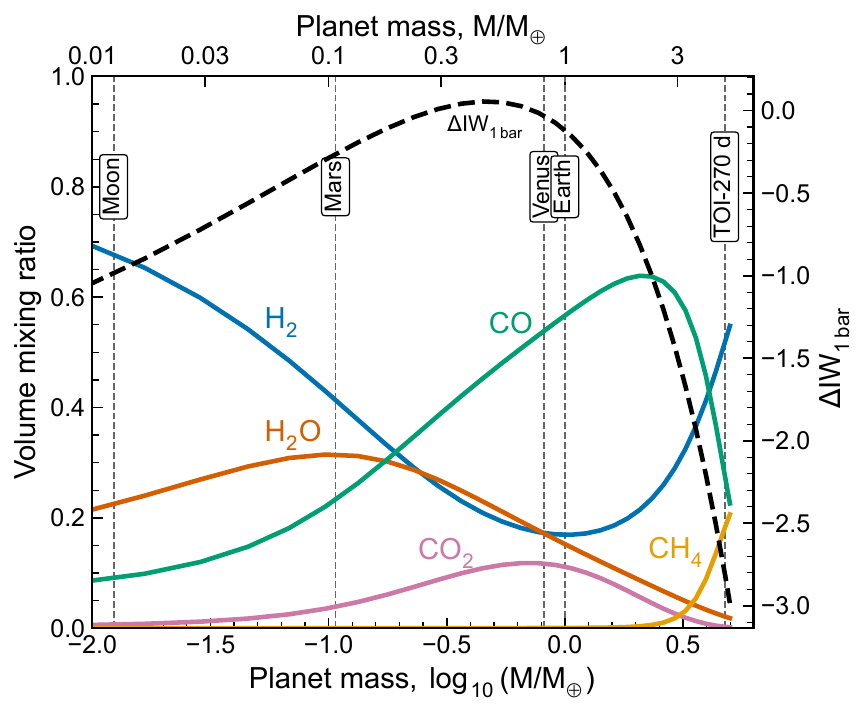}
    \caption{Computed volume mixing ratios of gas species (coloured lines) in the C-O-H system as a function of planet mass relative to that of Earth ($M/M_{\oplus}$), expressed as log$_{10}M/M_{\oplus}$. Concentrations of H and C were fixed at 0.06~wt.~\% and 0.02~wt.~\% of the planet mass, respectively. The dashed black curve corresponds to the oxygen fugacity at the surface of the planet at 2000~K relative to the iron-wüstite buffer ($\Delta$IW) at 1 bar \citep{hirschmann_iron-wustite_2021}. Vertical dashed lines denote the masses of a range of planetary bodies. See text for details. }
    \label{fig:atmosphere_vs_mass}
\end{figure}

To determine the chemical compositions of atmospheres in equilibrium with silicate magma oceans \texttt{atmodeller} \citep{bower2025} is used. Planet mass is varied between 10$^{-6}$ $<$ M/M$_{\oplus}$ $<$ 5, and $\Delta$ IW is calculated according to an empirical model that matches the $f$O$_2$s of the Moon, Mars and Earth and accounts for the spin transition at higher pressures (section \ref{sec:g_spin}). Temperature is kept constant at 2000~K and the H and C concentrations are fixed at 0.06~wt.~\% and 0.02~wt.~\% of the planet mass, respectively. Solubility relationships for each of the gas species considered H$_2$ \citep{hirschmann2012solubility}, H$_2$O \citep{sossi2023solubility}, CO \citep{yoshioka2019carbon}, CO$_2$ \citep{dixon1995} and CH$_4$ \citep{ardia2013solubility} are employed to compute their distribution between atmosphere and magma ocean, assuming the mantle mass fraction and bulk density are constant at 0.68 and 5500 kg/m$^3$, respectively. The volume mixing ratios ($p_i/P_T$) are shown in Fig.~\ref{fig:atmosphere_vs_mass}.

Striking is the gradual transition from H$_2$-rich atmospheres around small bodies such as the Moon, ceding to CO-rich compositions at the $f$O$_2$ maximum near Venus and Earth, before transitioning to H$_2$-, CH$_4$-rich varieties around super-Earths and sub-Neptunes. The differences in atmospheric mixing ratios at the same $\Delta$IW as a function of planet mass arises due to the effect of gravity on pressure, given by $P_a = M_ag/4 \pi r^2$, for a single-species atmosphere, $a$, of mass $M_a$ on a planet with radius $r$, which, in turn, influences homogeneous equilibria in the gas phase. In the system C-O-H, this is rationalised in the `Sabatier' reaction
\begin{equation}
    \mathrm{3H_2}(g) + \mathrm{CO}(g) = \mathrm{CH_4}(g) + \mathrm{H_2O}(g)
\end{equation}
which, assuming ideality, shows $p$CH$_4$/$p$CO $\propto$ ($p$H$_2$)$^3$. Indeed, methane appears to be the dominant gas, aside from hydrogen, in sub-Neptune atmospheres, such as TOI-270 d with \ensuremath{4.86 \pm 0.50\,M/M_{\oplus}} \citep{felix2025,madhusudhan2025}, while CO is most consistent with the absorption feature at 4.5 $\mu$m in the atmosphere of the super-Earth 55-Cancri-e \citep{hu2024}. Nevertheless, direct comparisons with JWST spectra are not straightforward, as TOI-270 d has a photosphere with a temperature of 330--390~K, depending on albedo, and pressures near $\sim$1~mbar \citep{madhusudhan2025}, both markedly lower than shown in Fig. \ref{fig:atmosphere_vs_mass}. This caveat notwithstanding, the modest mixing ratio of CH$_4$ ($\sim$1--5~\%) and abundance of H$_2$ (likely $>$85~\%) in the atmosphere of TOI-270 d implies H/C ratios higher \citep[$\sim$1.5--3 by weight or $\sim$100--200$\times$ solar metallicity;][]{felix2025} than those employed in Fig.~\ref{fig:atmosphere_vs_mass} (H/C = 0.379 by weight for\ensuremath{ M/M_{\oplus}} = 5; which is lower than the bulk H/C ratio of 3 due to the solubility of H$_2$O in the mantle). By contrast, since 55-Cancri-e has \ensuremath{M/M_{\oplus}} = 7.99\ensuremath{\pm}0.33, the fact it does not harbour an H$_2$-, CH$_4$-rich atmosphere implies either more oxidising conditions or lower H/C ratios than used to construct Fig.~\ref{fig:atmosphere_vs_mass}. 

The prediction that Fe$^{3+}$/$\sum$Fe reaches a maximum for Earth-like planets leads to CO-rich atmospheres. These occur because, at $\Delta$IW close to 0, $p$H$_2$/$p$H$_2$O is roughly unity, and water is far more soluble than H$_2$, meaning that much of the H budget of Earth-like planets is stored in their interiors. This property, combined with their higher mean molecular weight atmospheres, would have both acted to shield H from atmospheric escape on Earth-like planets. These atmospheres become CO$_2$-rich upon cooling \citep{sossi_redox_2020}, leading to conditions that might have been conducive to generating and sustaining habitable states on their surfaces.

%\noindent The foregoing analysis neglects the difference in composition between the glasses analysed by \citet{zhang_ferric_2024} and that of Earth's mantle. In particular, their glasses contain $\sim$20 -- 27~wt.~\% Al$_2$O$_3$, compared to 4.5~wt.~\% in the Earth's primitive mantle \citep{palme_cosmochemical_2013}. Owing to the favourable enthalpy change of coupled substitutions involving Al$^{3+}$Fe$^{3+}$ $\Leftrightarrow$ Mg$^{2+}$Si$^{4+}$ in bridgmanite \citep{frost2004}, we speculate that such Al-rich compositions may stabilise Fe$^{3+}$ moieities in high pressure silicate liquids with respect to peridotitic bulk compositions, which would therefore result in a further decrease in the Fe$^{3+}$/$\Sigma$Fe at a given $P-T-f$O$_2$ relative to the values given in \citet{zhang_ferric_2024}. 

%A larger influence of the spin transition on the stability of Fe$^{2+}$ relative to Fe$^{3+}$ is anticipated, owing to its greater propensity to exist in VI-fold coordination in silicate glasses and melts \citep{mysen_silicate_2018}. The non-negligible fraction (0.06 -- 0.09) of Fe$^{2+}_{LS}$ that is proposed to exist at 1 bar, therefore presumably reflects the proportion of Fe$^{2+}$ that existed in near-perfect octahedral coordination across the glass transition temperature. 

%%%%%%%%%%%%%%%%%%%%%%%%%%%%%%%%%%%%%%%%%%%%%%%%%%%%%%%%%%%%%%%%%%%%%%%%%%%%%%%%%%%%%%

\section{Conclusions}

Two sets of glasses were synthesised at low and high oxygen fugacities (\ensuremath{f\mathrm{O}_2}) with \ensuremath{\mathrm{Fe}^{3+}/\Sigma \mathrm{Fe} = 0.02(2)} and \ensuremath{1.00(2)}, respectively, determined by colourimetry. The glasses were compressed at room temperature up to 174~GPa in a diamond-anvil cell and \textit{in-situ} energy-domain synchrotron Mössbauer spectra were collected to determine the electronic state of iron. The ferric-iron-bearing glass show little change in the hyperfine parameters of the doublets with pressure, whereas those for their ferrous counterparts have two doublets that characterise the glass at 1~bar, \ensuremath{D_1} and \ensuremath{D_2}. With increasing pressure, the \ensuremath{D_2/(D_1 + D_2)} ratio increases to 0.65 by 172~GPa without any formation of \ensuremath{\mathrm{Fe}^{0}}, meaning that \ensuremath{D_2} is related to low-spin \ensuremath{\mathrm{Fe}^{2+}} and not \ensuremath{\mathrm{Fe}^{3+}}. This demonstrates that mid-ocean-ridge basaltic glasses have \ensuremath{\sim 10\%~\mathrm{Fe}^{3+}}, in agreement with wet chemical determinations. The spin transition in ferrous iron results in \ensuremath{\mathrm{Fe}^{3+}/\Sigma \mathrm{Fe}} ratios that decrease with increasing pressure beyond \ensuremath{\sim 40}~GPa. This pressure coincides with that inferred for core-mantle equilibrium on Earth, providing a mechanism to account for its uniquely high \ensuremath{\mathrm{Fe}^{3+}/\Sigma \mathrm{Fe}} among the terrestrial planets. By contrast, both smaller and larger planets are expected to be more reducing, resulting in secondary atmospheres that transition from \ensuremath{\mathrm{H}_2}-rich to CO-rich to \ensuremath{\mathrm{H}_2}- and \ensuremath{\mathrm{CH}_4}-rich for Moon-sized, Earth-sized, and super-Earth-like planets, respectively.

%%%%%%%%%%%%%%%%%%%%%%%%%%%%%%%%%%%%%%%%%%%%%%%%%%%%%%%%%%%%%%%%%%%%%%%%%%%%%%%%%%%%%%%%%%%%%%%%%%%%
%%%%%%%%%%%%%%%%%%%%%%%%%%%%%%%%%%%%%%%%%%%%%%%%%%%%%%%%%%%%%%%%%%%%%%%%%%%%%%%%%%%%%%%%%%%%%%%%%%%%
% Acknowledgments (not included in the word count)

\section*{Acknowledgments}
\noindent We thank two anonymous reviewers and the editor, Carolina Lithgow-Bertelloni, for perceptive comments on an earlier version of this work. We acknowledge the European Synchrotron Radiation Facility, Grenoble, France, for the provision of synchrotron radiation beamtime at the ID18 beamline (ES-959, ES-1252, ES-1328). We thank M. Murakami for lab resources, X. Li for assistance in Mössbauer spectroscopy measurements at ID18, C. La for assistance during colorimetry analysis, and J. Allaz for the technical assistance during electron microprobe analysis at ETH Zurich. \textbf{Funding}: This project was funded by the Swiss National Science Foundation (SNSF) grant No.197344 to Sylvain Petitgirard. PAS was supported by an SNSF Eccellenza Professorship (\# 203668) and the Swiss State Secretariat for Education, Research and Innovation (SERI) under contract No. MB22.00033, a SERI-funded ERC Starting grant “2ATMO”. 

\section*{Data availability}
\noindent All data needed to evaluate the conclusions in the paper are present in the paper and/or the Supplementary Materials and/or on the OSF Repository at \url{https://osf.io/kcrb9}.  

\section*{Competing interests} 
\noindent The authors declare that they have no competing interests.

\section*{CRediT authorship contribution statement}
\noindent \textbf{Alice Girani}: investigation, methodology, data curation, formal analysis, visualization, funding acquisition, Writing—original draft and writing—review and editing. \textbf{Sylvain Petitgirard}.: conceptualization, investigation, resources, methodology, funding acquisition, supervision, visualization, writing—review and editing, validation and project administration. \textbf{Sergey A. Yaroslavtsev}: investigation, methodology, writing—review and editing. \textbf{Giorgios Aprilis}: investigation, writing—review and editing. \textbf{James Badro}: investigation, writing—review and editing.\textbf{ Antoine Bézos}: investigation, writing—review and editing. \textbf{Hugh St.C. O'Neill}: investigation, writing—review and editing. \textbf{Paolo A. Sossi}: conceptualization, methodology, investigation, supervision, funding acquisition, visualization, validation, writing—review and editing and project administration. 

\clearpage
%%%%%%%%%%%%%%%%%%%%%%%%%%%%%%%%%%%%%%%%%%%%%%%%%%%%%%%%%%%%%%%%%%%%%%%%%%%%%%%%%%%%%%%--------------------------------------------------------------------------

\clearpage

%%%%%%%%%%%%%%%%%%%%%%%%%%%%%%%%%%%%%%%%%%%%%%%%%%%%%%%%%%%%%%%%%%%%%%%%%%%%%%%%%%%%%%%%%%%%%%%%%%%%

%% Bibliography
\bibliographystyle{elsarticle-harv} 
\bibliography{Bibliography}

@article{ardia2013solubility,
  title={Solubility of CH4 in a synthetic basaltic melt, with applications to atmosphere--magma ocean--core partitioning of volatiles and to the evolution of the Martian atmosphere},
  author={Ardia, P and Hirschmann, Marc M and Withers, AC and Stanley, BD},
  journal={Geochimica et Cosmochimica Acta},
  volume={114},
  pages={52--71},
  year={2013},
  publisher={Elsevier}
}

@inproceedings{asimow2012shock,
  title={Shock compression of preheated silicate liquids: Apparent universality of increasing Gr{\"u}neisen parameter upon compression},
  author={Asimow, Paul D},
  booktitle={AIP Conference Proceedings},
  volume={1426},
  number={1},
  pages={887--890},
  year={2012},
  organization={American Institute of Physics},
  doi={10.1063/1.3686420}
}

@article{berry2021,
  title={The coordination of Cr2+ in silicate glasses and implications for mineral-melt fractionation of Cr isotopes},
  author={Berry, Andrew J and Miller, Laura A and O'Neill, Hugh St C and Foran, Garry J},
  journal={Chemical Geology},
  volume={586},
  pages={120483},
  year={2021},
  publisher={Elsevier},
  doi={10.1016/j.chemgeo.2021.120483}
}

@article{bower2025,
  title={Diversity of Low-mass Planet Atmospheres in the C--H--O--N--S--Cl System with Interior Dissolution, Nonideality, and Condensation: Application to TRAPPIST-1e and Sub-Neptunes},
  author={Bower, Dan J and Thompson, Maggie A and Hakim, Kaustubh and Tian, Meng and Sossi, Paolo A},
  journal={The Astrophysical Journal},
  volume={995},
  number={1},
  pages={59},
  year={2025},
  publisher={The American Astronomical Society}
}

@article{desmarais2021,
  title={57Fe M{\"o}ssbauer isomer shift of pure iron and iron oxides at high pressure—An experimental and theoretical study},
  author={Desmarais, Jacques K and Bi, Wenli and Zhao, Jiyong and Hu, Michael Y and Alp, Esen and Tse, John S},
  journal={The Journal of Chemical Physics},
  volume={154},
  number={21},
  year={2021},
  publisher={AIP Publishing}
}

@article{dixon1995,
  title={An experimental study of water and carbon dioxide solubilities in mid-ocean ridge basaltic liquids. Part I: calibration and solubility models},
  author={Dixon, Jacqueline Eaby and Stolper, Edward M and Holloway, John R},
  journal={Journal of Petrology},
  volume={36},
  number={6},
  pages={1607--1631},
  year={1995},
  publisher={Oxford University Press}
}

@article{felix2025,
  title={Competing chemical signatures in the atmosphere of TOI-270 d: Inference of sulfur and carbon chemistry},
  author={Felix, Lukas and Kitzmann, Daniel and Demory, B-O and Mordasini, Christoph},
  journal={Astronomy \& Astrophysics},
  volume={701},
  pages={A296},
  year={2025},
  publisher={EDP Sciences}
}

@article{hirschmann2012solubility,
  title={Solubility of molecular hydrogen in silicate melts and consequences for volatile evolution of terrestrial planets},
  author={Hirschmann, Marc M and Withers, Anthony C and Ardia, Paola and Foley, NT},
  journal={Earth and Planetary Science Letters},
  volume={345},
  pages={38--48},
  year={2012},
  publisher={Elsevier}
}

@article{hu2024,
  title={A secondary atmosphere on the rocky exoplanet 55 Cancri e},
  author={Hu, Renyu and Bello-Arufe, Aaron and Zhang, Michael and Paragas, Kimberly and Zilinskas, Mantas and van Buchem, Christiaan and Bess, Michael and Patel, Jayshil and Ito, Yuichi and Damiano, Mario and others},
  journal={Nature},
  volume={630},
  number={8017},
  pages={609--612},
  year={2024},
  publisher={Nature Publishing Group UK London}
}

@article{kupenko2015oxidation,
  title={Oxidation state of the lower mantle: In situ observations of the iron electronic configuration in bridgmanite at extreme conditions},
  author={Kupenko, Ilya and McCammon, Catherine and Sinmyo, Ryosuke and Cerantola, Valerio and Potapkin, Vasily and Chumakov, Aleksandr I and Kantor, Anastasia and R{\"u}ffer, Rudolf and Dubrovinsky, Leonid},
  journal={Earth and Planetary Science Letters},
  volume={423},
  pages={78--86},
  year={2015},
  publisher={Elsevier},
  doi={10.1016/j.epsl.2015.04.027}
}

@article{madhusudhan2025,
  title={Exploring the sub-Neptune frontier with JWST},
  author={Madhusudhan, Nikku and Holmberg, M{\aa}ns and Constantinou, Savvas and Cooke, Gregory J},
  journal={Proceedings of the National Academy of Sciences},
  volume={122},
  number={39},
  pages={e2416194122},
  year={2025},
  publisher={National Academy of Sciences}
}

@article{mao2014,
  title={Spin and valence states of iron in Al-bearing silicate glass at high pressures studied by synchrotron M{\"o}ssbauer and X-ray emission spectroscopy},
  author={Mao, Zhu and Lin, Jung-Fu and Yang, Jing and Wu, Junjie and Watson, Heather C and Xiao, Yuming and Chow, Paul and Zhao, Jiyong},
  journal={American Mineralogist},
  volume={99},
  number={2-3},
  pages={415--423},
  year={2014},
  publisher={Mineralogical Society of America}
}

@article{mashino2025valence,
  title={Valence/spin states of iron in peridotite glass to megabar pressure implications for dense iron-rich silicate melt at the bottom of the mantle},
  author={Mashino, Izumi and Yoshino, Takashi and Mitsui, Takaya and Fujiwara, Kosuke and Inou{\'e}, Sayako and Sakai, Takeshi},
  journal={Geophysical Research Letters},
  volume={52},
  number={7},
  pages={e2024GL113106},
  year={2025},
  publisher={Wiley Online Library},
  doi={10.1029/2024GL113106}
}

@article{oneill2018oxidation,
  title={The oxidation state of iron in Mid-Ocean Ridge Basaltic (MORB) glasses: Implications for their petrogenesis and oxygen fugacities},
  author={O'Neill, Hugh St C and Berry, Andrew J and Mallmann, Guilherme},
  journal={Earth and Planetary Science Letters},
  volume={504},
  pages={152--162},
  year={2018},
  publisher={Elsevier},
  doi={10.1016/j.epsl.2018.10.002}
}

@article{oneill2024,
  title={The relationship between iron redox states and H2O contents in back-arc basin basaltic glasses from the North Fiji Basin},
  author={O'Neill, Hugh St C and Berry, Andrew J and Danyushevsky, Leonid V and Falloon, Trevor J and Maas, Roland and Feig, Sandrin T},
  journal={Chemical Geology},
  volume={655},
  pages={122062},
  year={2024},
  publisher={Elsevier},
  doi={10.1016/j.chemgeo.2024.122062}
}

@article{sinmyo2017spin,
  title={The spin state of Fe3+ in lower mantle bridgmanite},
  author={Sinmyo, Ryosuke and McCammon, Catherine and Dubrovinsky, Leonid},
  journal={American Mineralogist},
  volume={102},
  number={6},
  pages={1263--1269},
  year={2017},
  publisher={De Gruyter},
  doi={10.2138/am-2017-5917}
}

@article{siebert2012,
  title={Metal--silicate partitioning of Ni and Co in a deep magma ocean},
  author={Siebert, Julien and Badro, James and Antonangeli, Daniele and Ryerson, Frederick J},
  journal={Earth and Planetary Science Letters},
  volume={321},
  pages={189--197},
  year={2012},
  publisher={Elsevier}
}

@article{sossi2023solubility,
  title={Solubility of water in peridotite liquids and the prevalence of steam atmospheres on rocky planets},
  author={Sossi, Paolo A and Tollan, Peter ME and Badro, James and Bower, Dan J},
  journal={Earth and Planetary Science Letters},
  volume={601},
  pages={117894},
  year={2023},
  publisher={Elsevier}
}

@article{yaroslavtsev2025,
  title={Different widths of resonant lines in the M{\"o}ssbuaer spectrum as the result of multidimensional Gaussian distribution},
  author={Yaroslavtsev, Sergey},
  journal={Nuclear Instruments and Methods in Physics Research Section B: Beam Interactions with Materials and Atoms},
  volume={563},
  pages={165669},
  year={2025},
  publisher={Elsevier}
}

@article{yoshioka2019carbon,
  title={Carbon solubility in silicate melts in equilibrium with a CO-CO2 gas phase and graphite},
  author={Yoshioka, Takahiro and Nakashima, Daisuke and Nakamura, Tomoki and Shcheka, Svyatoslav and Keppler, Hans},
  journal={Geochimica et cosmochimica acta},
  volume={259},
  pages={129--143},
  year={2019},
  publisher={Elsevier}
}

@article{berry_re-assessment_2018,
	title = {A re-assessment of the oxidation state of iron in {MORB} glasses},
	volume = {483},
	issn = {0012-821X},
	doi = {10.1016/j.epsl.2017.11.032},
	language = {en},
	urldate = {2023-03-29},
	journal = {Earth and Planetary Science Letters},
	author = {Berry, Andrew J. and Stewart, Glen A. and O'Neill, Hugh St. C. and Mallmann, Guilherme and Mosselmans, J. Fred W.},
	month = feb,
	year = {2018},
	pages = {114--123},
}

@article{bancroft1970mossbauer,
  title={A M{\"o}ssbauer study of structure and bonding in iron (II) low-spin compounds},
  author={Bancroft, GM and Mays, MJ and Prater, BE},
  journal={Journal of the Chemical Society A: Inorganic, Physical, Theoretical},
  pages={956--968},
  year={1970},
  publisher={Royal Society of Chemistry}
}

@article{bezos2021unraveling,
  title={Unraveling the confusion over the iron oxidation state in MORB glasses},
  author={B{\'e}zos, Antoine and Guivel, Christ{\`e}le and La, Carole and Fougeroux, Thibault and Humler, Eric},
  journal={Geochimica et Cosmochimica Acta},
  volume={293},
  pages={28--39},
  year={2021},
  publisher={Elsevier},
  doi={10.1016/j.gca.2020.10.004}
}

@article{prescher_iron_2014,
	title = {Iron spin state in silicate glass at high pressure: {Implications} for melts in the {Earth's} lower mantle},
	volume = {385},
	issn = {0012-821X},
	shorttitle = {Iron spin state in silicate glass at high pressure},
	doi = {10.1016/j.epsl.2013.10.040},
	language = {en},
	urldate = {2023-03-29},
	journal = {Earth and Planetary Science Letters},
	author = {Prescher, C. and Weigel, C. and McCammon, C. and Narygina, O. and Potapkin, V. and Kupenko, I. and Sinmyo, R. and Chumakov, A. I. and Dubrovinsky, L.},
	month = jan,
	year = {2014},
	pages = {130--136},
}

@article{deng_magma_2020,
	title = {A magma ocean origin to divergent redox evolutions of rocky planetary bodies and early atmospheres},
	volume = {11},
	copyright = {2020 The Author(s)},
	issn = {2041-1723},
	doi = {10.1038/s41467-020-15757-0},
	language = {en},
	number = {1},
	urldate = {2023-03-29},
	journal = {Nat Commun},
	author = {Deng, Jie and Du, Zhixue and Karki, Bijaya B. and Ghosh, Dipta B. and Lee, Kanani K. M.},
	month = apr,
	year = {2020},
	pages = {2007},
}

@article{komabayashi_high-temperature_2010,
	title = {High-temperature compression of ferropericlase and the effect of temperature on iron spin transition},
	volume = {297},
	issn = {0012821X},
	doi = {10.1016/j.epsl.2010.07.025},
	language = {en},
	number = {3-4},
	urldate = {2023-08-14},
	journal = {Earth and Planetary Science Letters},
	author = {Komabayashi, Tetsuya and Hirose, Kei and Nagaya, Yukio and Sugimura, Emiko and Ohishi, Yasuo},
	month = sep,
	year = {2010},
	pages = {691--699},
}

@article{lelosqsossi2023,
  title={Atomic structure and physical properties of peridotite glasses at 1 bar},
  author={Le Losq, Charles and Sossi, Paolo A},
  journal={Frontiers in Earth Science},
  volume={11},
  pages={1040750},
  year={2023},
  publisher={Frontiers Media SA},
  doi={10.3389/feart.2023.1040750}
}

@article{zhang_determination_2018,
	title = {Determination of {Fe3}+/Σ{Fe} of {XANES} basaltic glass standards by {Mössbauer} spectroscopy and its application to the oxidation state of iron in {MORB}},
	volume = {479},
	issn = {0009-2541},
	doi = {10.1016/j.chemgeo.2018.01.006},
	language = {en},
	urldate = {2023-04-18},
	journal = {Chemical Geology},
	author = {Zhang, Hongluo L. and Cottrell, Elizabeth and Solheid, Peat A. and Kelley, Katherine A. and Hirschmann, Marc M.},
	month = feb,
	year = {2018},
	pages = {166--175},
}

@article{sossi_redox_2020,
	title = {Redox state of {Earth}’s magma ocean and its {Venus}-like early atmosphere},
	volume = {6},
	issn = {2375-2548},
	doi = {10.1126/sciadv.abd1387},
	language = {en},
	number = {48},
	urldate = {2023-06-27},
	journal = {Sci. Adv.},
	author = {Sossi, Paolo A. and Burnham, Antony D. and Badro, James and Lanzirotti, Antonio and Newville, Matt and O'Neill, Hugh St.C.},
	month = nov,
	year = {2020},
	pages = {eabd1387},
}

@article{armstrong_deep_2019,
	title = {Deep magma ocean formation set the oxidation state of {Earth}’s mantle},
	volume = {365},
	doi = {10.1126/science.aax8376},
	number = {6456},
	urldate = {2024-05-02},
	journal = {Science},
	author = {Armstrong, Katherine and Frost, Daniel J. and McCammon, Catherine A. and Rubie, David C. and Boffa Ballaran, Tiziana},
	month = aug,
	year = {2019},
	pages = {903--906},
}

@article{oneill_thermodynamics_1985,
	title = {Thermodynamics of {Co3O4}: a possible electron spin unpairing transition in {Co3}+},
	volume = {12},
	issn = {1432-2021},
	shorttitle = {Thermodynamics of {Co3O4}},
	doi = {10.1007/BF00308208},
	language = {en},
	number = {3},
	urldate = {2024-08-22},
	journal = {Phys Chem Minerals},
	author = {O'Neill, H.St. C.},
	month = jul,
	year = {1985},
	pages = {149--154},
}

@article{zhang_ferric_2024,
	title = {Ferric iron stabilization at deep magma ocean conditions},
	volume = {10},
	doi = {10.1126/sciadv.adp1752},
	number = {42},
	urldate = {2024-11-12},
	journal = {Science Advances},
	author = {Zhang, Hongluo L. and Hirschmann, Marc M. and Lord, Oliver T. and Rosenthal, Anja and Yaroslavtsev, Sergey and Cottrell, Elizabeth and Chumakov, Alexandr I. and Walter, Michael J.},
	month = oct,
	year = {2024},
	pages = {eadp1752},
}

@article{hirschmann_magma_2022,
	title = {Magma oceans, iron and chromium redox, and the origin of comparatively oxidized planetary mantles},
	volume = {328},
	issn = {0016-7037},
	doi = {10.1016/j.gca.2022.04.005},
	urldate = {2024-11-12},
	journal = {Geochimica et Cosmochimica Acta},
	author = {Hirschmann, M. M.},
	month = jul,
	year = {2022},
	pages = {221--241},
}

@article{canup_origin_2001,
	title = {Origin of the {Moon} in a giant impact near the end of the {Earth}'s formation},
	volume = {412},
	copyright = {2001 Macmillan Magazines Ltd.},
	issn = {1476-4687},
	doi = {10.1038/35089010},
	language = {en},
	number = {6848},
	urldate = {2024-11-12},
	journal = {Nature},
	author = {Canup, Robin M. and Asphaug, Erik},
	month = aug,
	year = {2001},
	pages = {708--712},
}

@article{caracas_meltcrystal_2019,
	title = {Melt–crystal density crossover in a deep magma ocean},
	volume = {516},
	issn = {0012-821X},
	doi = {10.1016/j.epsl.2019.03.031},
	urldate = {2024-11-12},
	journal = {Earth and Planetary Science Letters},
	author = {Caracas, Razvan and Hirose, Kei and Nomura, Ryuichi and Ballmer, Maxim D.},
	month = jun,
	year = {2019},
	pages = {202--211},
}

@article{frost_redox_2008,
	title = {The redox state of the mantle during and just after core formation},
	volume = {366},
	doi = {10.1098/rsta.2008.0147},
	number = {1883},
	urldate = {2024-11-12},
	journal = {Philosophical Transactions of the Royal Society A: Mathematical, Physical and Engineering Sciences},
	author = {Frost, D.j and Mann, U and Asahara, Y and Rubie, D.c},
	month = sep,
	year = {2008},
	pages = {4315--4337},
}

@article{badro_experimental_2021,
	title = {Experimental investigation of elemental and isotopic evaporation processes by laser heating in an aerodynamic levitation furnace},
	volume = {353},
	issn = {1778-7025},
	doi = {10.5802/crgeos.56},
	language = {fr},
	number = {1},
	urldate = {2024-11-12},
	journal = {Comptes Rendus. Géoscience},
	author = {Badro, James and Sossi, Paolo A. and Deng, Zhengbin and Borensztajn, Stephan and Wehr, Nicolas and Ryerson, Frederick J.},
	year = {2021},
	pages = {101--114},
}

@article{berry_xanes_2003,
	title = {{XANES} calibrations for the oxidation state of iron in a silicate glass},
	volume = {88},
	copyright = {De Gruyter expressly reserves the right to use all content for commercial text and data mining within the meaning of Section 44b of the German Copyright Act.},
	issn = {1945-3027},
	doi = {10.2138/am-2003-0704},
	language = {en},
	number = {7},
	urldate = {2024-11-12},
	journal = {American Mineralogist},
	author = {Berry, Andrew J. and O'Neill, Hugh St C. and Jayasuriya, Kasthuri D. and Campbell, Stewart J. and Foran, Garry J.},
	month = jul,
	year = {2003},
	pages = {967--977},
}

@article{davis_composition_2009,
	title = {The composition of {KLB}-1 peridotite},
	volume = {94},
	copyright = {De Gruyter expressly reserves the right to use all content for commercial text and data mining within the meaning of Section 44b of the German Copyright Act.},
	issn = {1945-3027},
	doi = {10.2138/am.2009.2984},
	language = {en},
	number = {1},
	urldate = {2024-11-12},
	journal = {American Mineralogist},
	author = {Davis, Fred A. and Tangeman, Jean A. and Tenner, Travis J. and Hirschmann, Marc M.},
	month = jan,
	year = {2009},
	pages = {176--180},
}

@article{akahama_pressure_2006,
	title = {Pressure calibration of diamond anvil {Raman} gauge to {310GPa}},
	volume = {100},
	issn = {0021-8979},
	doi = {10.1063/1.2335683},
	number = {4},
	urldate = {2024-11-12},
	journal = {Journal of Applied Physics},
	author = {Akahama, Yuichi and Kawamura, Haruki},
	month = aug,
	year = {2006},
	pages = {043516},
}

@article{kantor_bx90_2012,
	title = {{BX90}: {A} new diamond anvil cell design for {X}-ray diffraction and optical measurements},
	volume = {83},
	issn = {0034-6748},
	shorttitle = {{BX90}},
	doi = {10.1063/1.4768541},
	number = {12},
	urldate = {2024-11-12},
	journal = {Review of Scientific Instruments},
	author = {Kantor, I. and Prakapenka, V. and Kantor, A. and Dera, P. and Kurnosov, A. and Sinogeikin, S. and Dubrovinskaia, N. and Dubrovinsky, L.},
	month = dec,
	year = {2012},
	pages = {125102},
}

@article{potapkin_57fe_2012,
	title = {The {57Fe} {Synchrotron} {Mössbauer} {Source} at the {ESRF}},
	volume = {19},
	issn = {0909-0495},
	doi = {10.1107/S0909049512015579},
	language = {en},
	number = {4},
	urldate = {2024-11-12},
	journal = {J Synchrotron Rad},
	author = {Potapkin, V. and Chumakov, A. I. and Smirnov, G. V. and Celse, J.-P. and Rüffer, R. and McCammon, C. and Dubrovinsky, L.},
	month = jul,
	year = {2012},
	pages = {559--569},
}

@article{virgo_structural_1985,
	title = {The structural state of iron in oxidized vs. reduced glasses at 1 atm: {A57Fe} {Mössbauer} study},
	volume = {12},
	issn = {1432-2021},
	shorttitle = {The structural state of iron in oxidized vs. reduced glasses at 1 atm},
	doi = {10.1007/BF01046829},
	language = {en},
	number = {2},
	urldate = {2024-11-12},
	journal = {Phys Chem Minerals},
	author = {Virgo, David and Mysen, Bjørn O.},
	month = mar,
	year = {1985},
	pages = {65--76},
}

@article{yaroslavtsev_syncmoss_2023,
	title = {{SYNCmoss} software package for fitting {Mössbauer} spectra measured with a synchrotron {Mössbauer} source},
	volume = {30},
	copyright = {https://creativecommons.org/licenses/by/4.0/},
	issn = {1600-5775},
	doi = {10.1107/S1600577523001686},
	language = {en},
	number = {3},
	urldate = {2024-11-12},
	journal = {J Synchrotron Rad},
	author = {Yaroslavtsev, S.},
	month = may,
	year = {2023},
	pages = {596--604},
}

@article{murakami_high-pressure_2014,
	title = {High-pressure radiative conductivity of dense silicate glasses with potential implications for dark magmas},
	volume = {5},
	copyright = {2014 Springer Nature Limited},
	issn = {2041-1723},
	doi = {10.1038/ncomms6428},
	language = {en},
	number = {1},
	urldate = {2024-11-12},
	journal = {Nat Commun},
	author = {Murakami, Motohiko and Goncharov, Alexander F. and Hirao, Naohisa and Masuda, Ryo and Mitsui, Takaya and Thomas, Sylvia-Monique and Bina, Craig R.},
	month = nov,
	year = {2014},
	pages = {5428},
}

@article{maeda_spin_2017,
	title = {Spin state and electronic environment of iron in basaltic glass in the lower mantle},
	volume = {102},
	copyright = {De Gruyter expressly reserves the right to use all content for commercial text and data mining within the meaning of Section 44b of the German Copyright Act.},
	issn = {1945-3027},
	doi = {10.2138/am-2017-6035},
	language = {en},
	number = {10},
	urldate = {2024-11-12},
	journal = {American Mineralogist},
	author = {Maeda, Fumiya and Kamada, Seiji and Ohtani, Eiji and Hirao, Naohisa and Mitsui, Takaya and Masuda, Ryo and Miyahara, Masaaki and McCammon, Catherine},
	month = oct,
	year = {2017},
	pages = {2106--2112},
}

@article{solomatova_electronic_2017,
	title = {Electronic environments of ferrous iron in rhyolitic and basaltic glasses at high pressure},
	volume = {122},
	copyright = {©2017. American Geophysical Union. All Rights Reserved.},
	issn = {2169-9356},
	doi = {10.1002/2017JB014363},
	language = {en},
	number = {8},
	urldate = {2024-11-12},
	journal = {Journal of Geophysical Research: Solid Earth},
	author = {Solomatova, Natalia V. and Jackson, Jennifer M. and Sturhahn, Wolfgang and Rossman, George R. and Roskosz, Mathieu},
	year = {2017},
	pages = {6306--6322},
}

@article{jayasuriya_mossbauer_2004,
	title = {A {Mössbauer} study of the oxidation state of {Fe} in silicate melts},
	volume = {89},
	copyright = {De Gruyter expressly reserves the right to use all content for commercial text and data mining within the meaning of Section 44b of the German Copyright Act.},
	issn = {1945-3027},
	doi = {10.2138/am-2004-11-1203},
	language = {en},
	number = {11-12},
	urldate = {2024-11-12},
	journal = {American Mineralogist},
	author = {Jayasuriya, Kasthuri D. and O'Neill, Hugh St C. and Berry, Andrew J. and Campbell, Stewart J.},
	month = nov,
	year = {2004},
	pages = {1597--1609},
}

@incollection{mccammon_mossbauer_2021,
	address = {Singapore},
	title = {Mössbauer {Spectroscopy} with {High} {Spatial} {Resolution}: {Spotlight} on {Geoscience}},
	isbn = {9789811594229},
	shorttitle = {Mössbauer {Spectroscopy} with {High} {Spatial} {Resolution}},
	language = {en},
	urldate = {2024-11-12},
	booktitle = {Modern {Mössbauer} {Spectroscopy}: {New} {Challenges} {Based} on {Cutting}-{Edge} {Techniques}},
	publisher = {Springer},
	author = {McCammon, Catherine},
	editor = {Yoshida, Yutaka and Langouche, Guido},
	year = {2021},
	doi = {10.1007/978-981-15-9422-9_5},
	pages = {221--266},
}

@article{mccammon_iron_2013,
	title = {Iron spin state in silicate perovskite at conditions of the {Earth}'s deep interior},
	volume = {33},
	issn = {0895-7959},
	doi = {10.1080/08957959.2013.805217},
	number = {3},
	urldate = {2024-11-12},
	journal = {High Pressure Research},
	author = {McCammon, Catherine and Glazyrin, Konstantin and Kantor, Anastasia and Kantor, Innokenty and Kupenko, Ilya and Narygina, Olga and Potapkin, Vasily and Prescher, Clemens and Sinmyo, Ryosuke and Chumakov, Alexandr and Rüffer, Rudolf and Sergueev, Ilya and Smirnov, Gennady and Dubrovinsky, Leonid},
	month = aug,
	year = {2013},
	pages = {663--672},
}

@article{shim_ultrafast_2023,
	title = {Ultrafast x-ray detection of low-spin iron in molten silicate under deep planetary interior conditions},
	volume = {9},
	doi = {10.1126/sciadv.adi6153},
	number = {42},
	urldate = {2024-11-12},
	journal = {Science Advances},
	author = {Shim, Sang-Heon and Ko, Byeongkwan and Sokaras, Dimosthenis and Nagler, Bob and Lee, He Ja and Galtier, Eric and Glenzer, Siegfried and Granados, Eduardo and Vinci, Tommaso and Fiquet, Guillaume and Dolinschi, Jonathan and Tappan, Jackie and Kulka, Britany and Mao, Wendy L. and Morard, Guillaume and Ravasio, Alessandra and Gleason, Arianna and Alonso-Mori, Roberto},
	month = oct,
	year = {2023},
	pages = {eadi6153},
}

@article{speziale_iron_2005,
	title = {Iron spin transition in {Earth}'s mantle},
	volume = {102},
	doi = {10.1073/pnas.0508919102},
	number = {50},
	urldate = {2024-11-12},
	journal = {Proceedings of the National Academy of Sciences},
	author = {Speziale, S. and Milner, A. and Lee, V. E. and Clark, S. M. and Pasternak, M. P. and Jeanloz, R.},
	month = dec,
	year = {2005},
	pages = {17918--17922},
}

@article{rouquette_high-pressure_2008,
	title = {High-{Pressure} {Studies} of ({Mg0}.{9Fe0}.1){2SiO4} {Olivine} {Using} {Raman} {Spectroscopy}, {X}-ray {Diffraction}, and {Mössbauer} {Spectroscopy}},
	volume = {47},
	issn = {0020-1669},
	doi = {10.1021/ic701983w},
	number = {7},
	urldate = {2024-11-12},
	journal = {Inorg. Chem.},
	author = {Rouquette, J. and Kantor, I. and McCammon, C. A. and Dmitriev, V. and Dubrovinsky, L. S.},
	month = apr,
	year = {2008},
	pages = {2668--2673},
}

@article{finklea_investigation_1976,
	title = {Investigation of the bonding mechanism in pyrite using the {Mössbauer} effect and {X}-ray crystallography},
	volume = {32},
	issn = {0567-7394},
	doi = {10.1107/S0567739476001198},
	language = {en},
	number = {4},
	urldate = {2024-11-12},
	journal = {Acta Cryst A},
	author = {Finklea, S. L. and Cathey, L. and Amma, E. L.},
	month = jul,
	year = {1976},
	pages = {529--537},
}

@article{cottrell_high-precision_2009,
	title = {High-precision determination of iron oxidation state in silicate glasses using {XANES}},
	volume = {268},
	issn = {0009-2541},
	doi = {10.1016/j.chemgeo.2009.08.008},
	number = {3},
	urldate = {2024-11-13},
	journal = {Chemical Geology},
	author = {Cottrell, Elizabeth and Kelley, Katherine A. and Lanzirotti, Antonio and Fischer, Rebecca A.},
	month = nov,
	year = {2009},
	pages = {167--179},
}

@article{cottrell_oxidation_2011,
	title = {The oxidation state of {Fe} in {MORB} glasses and the oxygen fugacity of the upper mantle},
	volume = {305},
	issn = {0012-821X},
	doi = {10.1016/j.epsl.2011.03.014},
	number = {3},
	urldate = {2024-11-13},
	journal = {Earth and Planetary Science Letters},
	author = {Cottrell, Elizabeth and Kelley, Katherine A.},
	month = may,
	year = {2011},
	pages = {270--282},
}

@article{hirschmann_iron-wustite_2021,
	title = {Iron-wüstite revisited: {A} revised calibration accounting for variable stoichiometry and the effects of pressure},
	volume = {313},
	issn = {0016-7037},
	shorttitle = {Iron-wüstite revisited},
	doi = {10.1016/j.gca.2021.08.039},
	urldate = {2024-11-13},
	journal = {Geochimica et Cosmochimica Acta},
	author = {Hirschmann, M. M.},
	month = nov,
	year = {2021},
	pages = {74--84},
}

@article{belonoshko_molecular_1991,
	title = {A molecular dynamics study of the pressure-volume-temperature properties of super-critical fluids: {I}. {H2O}},
	volume = {55},
	issn = {0016-7037},
	shorttitle = {A molecular dynamics study of the pressure-volume-temperature properties of super-critical fluids},
	doi = {10.1016/0016-7037(91)90425-5},
	number = {1},
	urldate = {2024-11-13},
	journal = {Geochimica et Cosmochimica Acta},
	author = {Belonoshko, A and Saxena, S. K},
	month = jan,
	year = {1991},
	pages = {381--387},
}

@article{alberto_analysis_1996,
	title = {Analysis of {Mössbauer} spectra of silicate glasses using a two-dimensional {Gaussian} distribution of hyperfine parameters},
	volume = {194},
	issn = {0022-3093},
	doi = {10.1016/0022-3093(95)00463-7},
	number = {1},
	urldate = {2024-11-13},
	journal = {Journal of Non-Crystalline Solids},
	author = {Alberto, H. V. and Pinto da Cunha, J. L. and Mysen, B. O. and Gil, J. M. and Ayres de Campos, N.},
	month = jan,
	year = {1996},
	pages = {48--57},
}

@article{brabers_low-spin-high-spin_1992,
	title = {Low-spin-high-spin transition in the {Co3O4} spinel},
	volume = {104-107},
	issn = {0304-8853},
	doi = {10.1016/0304-8853(92)90853-G},
	urldate = {2024-11-13},
	journal = {Journal of Magnetism and Magnetic Materials},
	author = {Brabers, V. A. M. and Broemme, A. D. D.},
	month = feb,
	year = {1992},
	pages = {405--406},
}

@article{carpentersaljie1994,
  title={Thermodynamics of nonconvergent cation ordering in minerals: I. An alternative approach},
  author={Carpenter, Michael A and Powell, Roger and Salje, Ekhard KH},
  journal={American Mineralogist},
  volume={79},
  number={11-12},
  pages={1053--1067},
  year={1994},
  publisher={Mineralogical Society of America}
}

@article{gilletrichet1997,
  title={Pressure-induced amorphization of minerals: a review},
  author={Richet, Pascal and Gillet, Philippe},
  journal={European Journal of Mineralogy-Ohne Beihefte},
  volume={9},
  number={5},
  pages={907--934},
  year={1997},
  publisher={Stuttgart: E. Schweizerbart'sche Verlagsbuchhandlung (Nagele u. Obermiller~…},
  doi={10.1127/ejm/9/5/0907}
}

@article{huang_structural_2022,
	title = {Structural evolution in a pyrolitic magma ocean under mantle conditions},
	volume = {584},
	issn = {0012-821X},
	doi = {10.1016/j.epsl.2022.117473},
	urldate = {2024-11-13},
	journal = {Earth and Planetary Science Letters},
	author = {Huang, Dongyang and Murakami, Motohiko and Brodholt, John and McCammon, Catherine and Petitgirard, Sylvain},
	month = apr,
	year = {2022},
	pages = {117473},
}

@book{robiehemingway1995,
  title={Thermodynamic properties of minerals and related substances at 298.15 K and 1 bar (10$^{5}$ Pascals) pressure and at higher temperatures},
  author={Robie, Richard A and Hemingway, Bruce S},
  volume={2131},
  year={1995},
  publisher={US Government Printing Office}
}

@article{roskosz_structural_2022,
	title = {Structural, redox and isotopic behaviors of iron in geological silicate glasses: {A} {NRIXS} study of {Lamb}-{Mössbauer} factors and force constants},
	volume = {321},
	issn = {0016-7037},
	shorttitle = {Structural, redox and isotopic behaviors of iron in geological silicate glasses},
	doi = {10.1016/j.gca.2022.01.021},
	urldate = {2024-11-13},
	journal = {Geochimica et Cosmochimica Acta},
	author = {Roskosz, Mathieu and Dauphas, Nicolas and Hu, Justin and Hu, Michael Y. and Neuville, Daniel R. and Brown, Dennis and Bi, Wenli and Nie, Nicole X. and Zhao, Jiyong and Alp, Esen E.},
	month = mar,
	year = {2022},
	pages = {184--205},
}

@article{cottrell2021,
  title={Oxygen fugacity across tectonic settings},
  author={Cottrell, Elizabeth and Birner, Suzanne K and Brounce, Maryjo and Davis, Fred A and Waters, Laura E and Kelley, Katherine A},
  journal={Magma redox geochemistry},
  pages={33--61},
  year={2021},
  publisher={Wiley Online Library},
  doi={10.1002/9781119473206.ch3}
}

@article{kuwahara2023hadean,
  title={Hadean mantle oxidation inferred from melting of peridotite under lower-mantle conditions},
  author={Kuwahara, Hideharu and Nakada, Ryoichi and Kadoya, Shintaro and Yoshino, Takashi and Irifune, Tetsuo},
  journal={Nature Geoscience},
  volume={16},
  number={5},
  pages={461--465},
  year={2023},
  publisher={Nature Publishing Group UK London},
  doi={10.1038/s41561-023-01169-4}
}

@article{gaillard_redox_2022,
	title = {Redox controls during magma ocean degassing},
	volume = {577},
	issn = {0012-821X},
	doi = {10.1016/j.epsl.2021.117255},
	urldate = {2024-11-21},
	journal = {Earth and Planetary Science Letters},
	author = {Gaillard, Fabrice and Bernadou, Fabien and Roskosz, Mathieu and Bouhifd, Mohamed Ali and Marrocchi, Yves and Iacono-Marziano, Giada and Moreira, Manuel and Scaillet, Bruno and Rogerie, Gregory},
	month = jan,
	year = {2022},
	pages = {117255},
}

@article{wilke2005,
  title={Fe in magma-an overview},
  author={Wilke, Max},
  journal={Annals of geophysics},
  volume={48},
  number={4-5},
  pages = {609 - 617},
  year={2005},
  doi={10.4401/ag-3222}
}

@article{wilson1960,
  title={The micro-determination of ferrous iron in silicate minerals by a volumetric and a colorimetric method},
  author={Wilson, AD},
  journal={Analyst},
  volume={85},
  number={1016},
  pages={823--827},
  year={1960},
  publisher={Royal Society of Chemistry},
  doi={10.1039/AN9608500823}
}

@article{nomura2011,
  title={Spin crossover and iron-rich silicate melt in the Earth’s deep mantle},
  author={Nomura, Ryuichi and Ozawa, Haruka and Tateno, Shigehiko and Hirose, Kei and Hernlund, John and Muto, Shunsuke and Ishii, Hirofumi and Hiraoka, Nozomu},
  journal={Nature},
  volume={473},
  number={7346},
  pages={199--202},
  year={2011},
  publisher={Nature Publishing Group UK London}
}

\end{document}